\documentclass[12pt]{article}

\usepackage{amsmath}
\usepackage{amssymb}
\usepackage{amsthm}
\usepackage{array}
\usepackage{bm}
\usepackage[footnotesize,bf]{caption}
\usepackage{color}
\usepackage{enumitem}
\usepackage{eurosym}
\usepackage[top=2.4cm,left=2.7cm,right=2.7cm,bottom=3cm]{geometry}
\usepackage{graphicx}
\usepackage{lineno}
\usepackage[sc]{mathpazo}
\usepackage{mathtools}
\usepackage{multirow}
\usepackage[numbers,sort&compress]{natbib}
\usepackage{setspace}
\usepackage{times}
\usepackage{titlesec}
\usepackage[normalem]{ulem}
\usepackage{verbatim}
\usepackage{xr}

\usepackage{url}
\usepackage{xcolor,graphicx,float}
\usepackage{color}
\usepackage{booktabs}
\usepackage{array}
\usepackage{changepage}

\newcommand*\patchAmsMathEnvironmentForLineno[1]{%
	\expandafter\let\csname old#1\expandafter\endcsname\csname #1\endcsname
	\expandafter\let\csname oldend#1\expandafter\endcsname\csname end#1\endcsname
	\renewenvironment{#1}%
	{\linenomath\csname old#1\endcsname}%
	{\csname oldend#1\endcsname\endlinenomath}}%
\newcommand*\patchBothAmsMathEnvironmentsForLineno[1]{%
	\patchAmsMathEnvironmentForLineno{#1}%
	\patchAmsMathEnvironmentForLineno{#1*}}%
\AtBeginDocument{%
	\patchBothAmsMathEnvironmentsForLineno{equation}%
	\patchBothAmsMathEnvironmentsForLineno{align}%
	\patchBothAmsMathEnvironmentsForLineno{flalign}%
	\patchBothAmsMathEnvironmentsForLineno{alignat}%
	\patchBothAmsMathEnvironmentsForLineno{gather}%
	\patchBothAmsMathEnvironmentsForLineno{multline}%
}

\smallskip

\definecolor{mygreen}{rgb}{0.0, 0.6, 0.0}

\titleformat{\section}{\sffamily \fontsize{13}{16}\bfseries}{\thesection}{1em}{}
\titleformat{\subsection}{\sffamily \fontsize{11.5}{11.5}\bfseries}{\thesubsection}{1em}{}

\theoremstyle{definition}

\definecolor{qired}{rgb}{0.6, 0, 0}

\title{\begin{center} \bfseries \singlespacing
\large
Evolution of cooperation with asymmetric social interactions
\end{center}
\date{}}
\author{\parbox[c]{16cm}{\onehalfspacing \centering ~\\[-0.4cm] Qi Su$^{1,2,3,}\footnote{Correspondence: qisu1991@sas.upenn.edu.}$  \quad Joshua B. Plotkin$^{1,2,3,}\footnote{Correspondence: jplotkin@sas.upenn.edu}$\\ \quad \\
		\footnotesize
		$^{1}$Department of Biology, University of Pennsylvania, Philadelphia, PA~19104 USA \\
		$^{2}$Center for Mathematical Biology, University of Pennsylvania, Philadelphia, PA~19104 USA \\
		$^{3}$Department of Mathematics, University of Pennsylvania, Philadelphia, PA~19104 USA \\[0.2cm]}
}

\begin{document}

\maketitle
\vspace*{-.2in}
\begin{center}
\date{May 20, 2021}   
\end{center}


\onehalfspacing

\begin{abstract}
%
\noindent How cooperation emerges in human societies is both an evolutionary enigma, and a practical problem with tangible implications for societal health. Population structure has long been recognized as a catalyst for cooperation because local interactions enable reciprocity. Analysis of this phenomenon typically assumes bi-directional social interactions, even though real-world interactions are often uni-directional. Uni-directional interactions -- where one individual has the opportunity to contribute altruistically to another, but not conversely -- arise in real-world populations as the result of organizational hierarchies, social stratification, popularity effects, and endogenous mechanisms of network growth. Here we expand the theory of cooperation in structured populations to account for both uni- and bi-directional social interactions. Even though directed interactions remove the opportunity for reciprocity, we find that cooperation can nonetheless be favored in directed social networks and that cooperation is provably maximized for networks with an intermediate proportion of directed interactions, as observed in many empirical settings.  We also identify two simple structural motifs that allow efficient modification of interaction directionality to promote cooperation by orders of magnitude. We discuss how our results relate to the concepts of generalized and indirect reciprocity.
\end{abstract}

\bigskip

\section{Introduction}

The past year has revealed a remarkable capacity for individuals to incur personal costs for the benefit of the common good as we confront a global pandemic -- such as volunteering as front-line workers, donating protective materials and supplies, and adhering to quarantine policies \cite{Bavel2020,Li2020b}. The question of why some societies, but not others, have readily met the challenge of costly cooperation for the common good reflects a classic problem in evolutionary theory: why would an individual forgo her own interests to help strangers? Understanding the spread and maintenance of cooperation is now recognized as an important practical problem, with tangible benefits. This is especially true today, as our globe tackles collective action problems in public health, resource management, and climate change \cite{1968-Hardin-p1243-1248}.

~\\
The last few decades have seen a proliferation of theoretical research into the evolutionary origins of cooperation and the dynamics of its spread. The literature has revealed several key insights into this enigma \cite{2006-Nowak-p1560-1563}.
Population structure is perhaps the most widely discussed mechanism that can promote cooperation \cite{1964-Hamilton-p1-16,1964-Hamilton-jtb2}, and it has been studied by computer simulation \cite{1992-Nowak-p826-829,Santos2005,Li2020}, mathematical analysis \cite{2005-Lieberman-p312-316,2006-Ohtsuki-p502-505,2006-Pacheco-p258103,2009-Tarnita-p570-581,2011-Hadjichrysanthou-p386-386,2013-Chen-p637-664,2017-Allen-p227-230,2019-Qi-p20190041-20190041,Allen2019,mcavoy-2020-nhb,2020-qi-arxiv,McAvoy2021}, and experiments with human subjects \cite{2014-Rand-p17093-17098}.
In structured populations individuals interact only with their neighbors -- through either physical or social ties -- and behaviors also spread locally. Population structure has the potential to favor the evolution of cooperative behaviors that would otherwise be disfavored in a well-mixed populations \cite{1964-Hamilton-p1-16,1964-Hamilton-jtb2,1992-Nowak-p826-829,2006-Ohtsuki-p502-505,2017-Allen-p227-230,2019-Qi-p20190041-20190041}.
In network-structured populations, for example, nodes represent individuals and edges typically represent social interactions between connected individuals \cite{2005-Lieberman-p312-316,2006-Ohtsuki-p502-505,2006-Pacheco-p258103,2011-Hadjichrysanthou-p386-386,2013-Chen-p637-664,2017-Allen-p227-230,2019-Qi-p20190041-20190041,Allen2019,mcavoy-2020-nhb,Li2020};
in set-structured populations, each individual is located in one or more of several social circles  \cite{2009-Tarnita-p8601-8604};
and in multilayer-structured populations, social interactions occur in multiple different domains, such as on-line and off-line interactions, and payoffs to an individual are summed across domains \cite{2015-Wang-p124-124,2020-qi-arxiv}.

~\\
Despite different approaches to describing population structure, nearly all research on this topic has assumed that social interactions and behavioral spread are bi-directional \cite{1992-Nowak-p826-829,Santos2005,Li2020,2005-Lieberman-p312-316,2006-Ohtsuki-p502-505,2006-Pacheco-p258103,2009-Tarnita-p570-581,2011-Hadjichrysanthou-p386-386,2013-Chen-p637-664,2017-Allen-p227-230,2019-Qi-p20190041-20190041,Allen2019,mcavoy-2020-nhb,2020-qi-arxiv,2014-Rand-p17093-17098}.
That is, if Charlie provides a benefit to Bob when behaving as a cooperator, Bob is presumed to provide a benefit to Charlie when acting as a cooperator; moreover, if Charlie has a chance to imitate Bob's behavior, then so too can Bob imitate Charlie's behavior.
The assumption of bi-directionality simplifies analysis and enables simple intuitions 
for why population structures permit the evolution of cooperation  \cite{2006-Ohtsuki-p502-505,Su2019pnas}. 

~\\
But bi-directional models neglect the prevalence of asymmetric social relationships in the real world \cite{Malliaros2013}.
For example, in an empirical network of inter-family relationships in San Juan Sur, a village in Costa Rica, almost two-thirds of the social ties are unidirectional: one family frequency visits another family, while the later family never reciprocates with a return visit \cite{nooy-2018}. And 
in the empirical friendship network of an Australian National University campus, more than a half of the relationships are unidirectional: one student regards another as her friend, but not conversely \cite{Freeman1998}. In the network of Twitter followers (based on a snowball sample crawl across “quality” users in 2009), more than 99\% of follower relationships are strictly unidirectional \cite{DeChoudhury2010}. 
Other examples include email networks \cite{Newman2002}, trust and advice consulting networks \cite{Coleman1957}, and social donations \cite{Burum2020} --- which all exhibit a high proportion of uni-directional social interactions.
Asymmetric interactions are also widespread outside of the human social domain, in systems ranging from international trade (trade volumes between countries \cite{Malliaros2013}) to rivers and stream flow (movement of microorganisms, nutrients and organic matters \cite{Brown2011,Liu2013,Mansour2018}).

~\\
Recent advances in network science have established that edge directionality can qualitatively alter dynamics across a range of systems \cite{2015-Pastor-Satorras-p925-979}, including in disease spread \cite{Meyers2006} and synchronization \cite{Restrepo2006}.
%
The empirical prevalence of directed social interactions, and its remarkable impact on dynamics in other settings, leaves an open question: how does directionality affect the evolution of cooperation?

~\\
Directed interactions are likely to fundamentally alter the evolution of cooperation, compared to the classic case of bi-directional interactions. Notably, bi-directionality allows for reciprocity. In the bi-directional setting, after an individual imitates her neighbor's behavior, the neighbor will then experience reciprocity -- so that two cooperative neighbors help each other, and two defecting neighbors harm each other. This phenomenon of ``network reciprocity" \cite{2006-Nowak-p1560-1563} along bi-directional edges is known to facilitate the local spread of cooperation and retard the spread of defection. But reciprocity cannot occur when behavioral imitation occurs along a uni-directional edge: when an individual copies her neighbor's strategy she does not have the opportunity to reciprocate the behavior she has imitated. Since reciprocity is disrupted, directionality may make it difficult, or even impossible, for cooperation to emerge in structured populations.

~\\
Here we study the evolution of cooperation in structured populations with uni-directional interactions. We uncover a surprising and general result: directionality can actually facilitate cooperation, even though it disrupts  reciprocity. We prove analytically that cooperation can evolve in populations with directional interactions, and that an intermediate level of directionality is most beneficial for cooperation. In fact, converting a portion of links to be uni-directional can even rescue cooperation on a bi-directional network whose topology otherwise prevents the emergence of cooperation. Analysis of four empirical social networks shows that the directionality measured in real-world settings facilitates the emergence of cooperation, compared to the outcome of bi-directional interactions along the same empirical network topology. Furthermore, we identify two simple network motifs that are critical to determining the evolution of cooperation, and which provide insights into how best to optimize edge directions to stimulate cooperation, by orders of magnitude. Our analysis reveals a profound effect of asymmetric social interactions for the evolution of behavior in structured populations.

\section{Model}
\begin{figure*}[!h]
\centering
\includegraphics[width=0.5\textwidth]{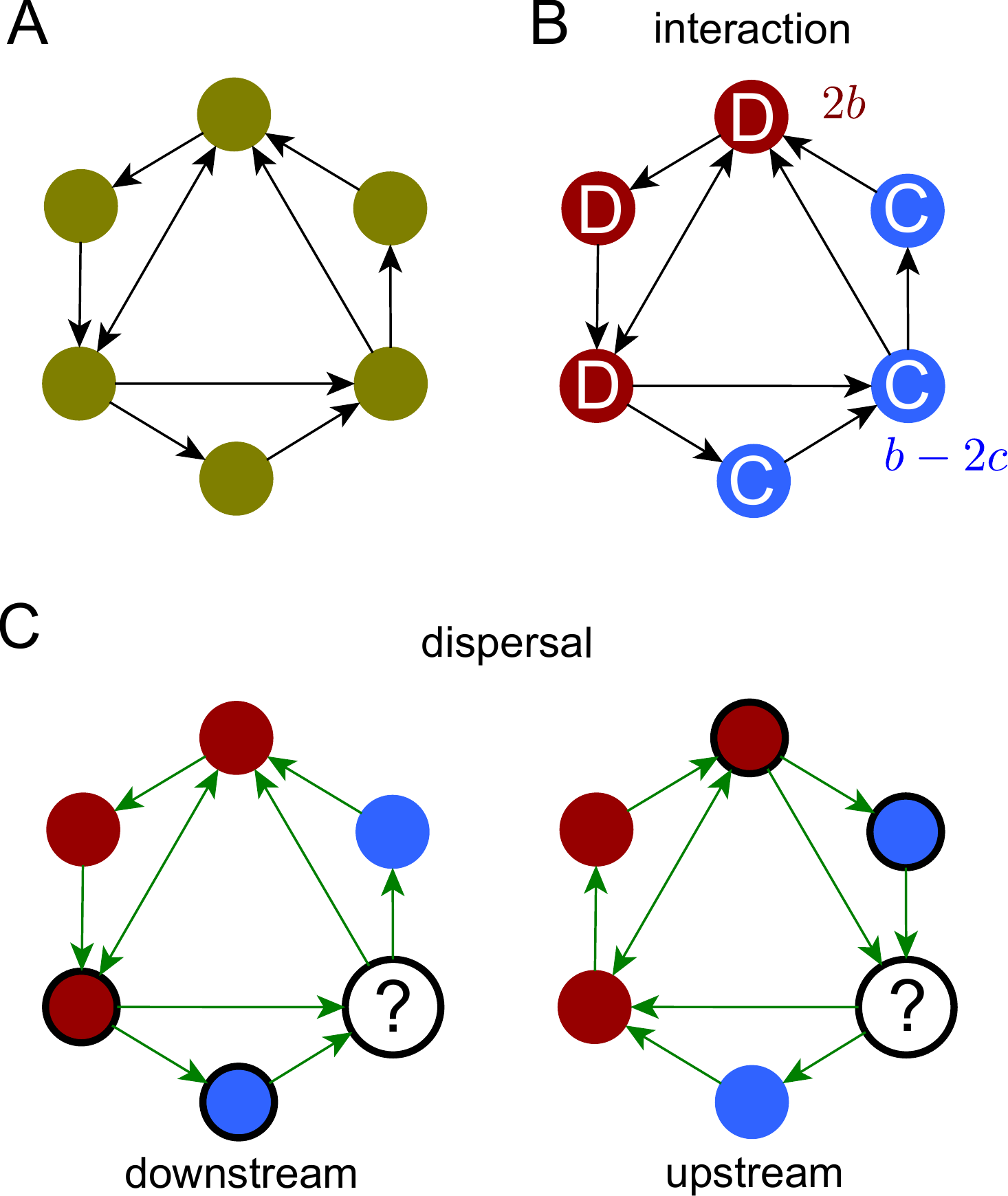}
\caption{\label{fig:1} \textbf{Evolutionary games on directed networks}.
(A) The population structure is described by a directed network. (B)
Individuals (nodes) engage in pairwise social interactions along edges. A cooperative individual ($C$) at a source node pays a cost $c$ to generate a benefit $b$ to a recipient at a target node; whereas a defector ($D$) pays no cost and generates no benefit.
Each payer $i$ accumulates a total payoff, summed over all pairwise interactions (two examples are indicated).
(C) After accumulating payoffs, a random individual $i$ (denoted by a question mark) is chosen to update her strategy according to payoff-biased imitation. All neighbors pointing to $i$ in the dispersal network compete to spread their strategy to $i$, with a probability proportional to their reproductive rate. We consider two cases of strategy dispersal: in the `downstream' case strategies disperse in the same direction as social interactions occur; in the `upstream' case strategies disperse in the opposite direction.
}
\end{figure*}

We model a population of $N$ individuals engaged in pairwise social interactions with their neighbors. Each player engages with each of her neighbors in a simple social dilemma called the donation game \cite{sigmund-2010}, choosing either to cooperate ($C$) or defect ($D$). A cooperative act means paying a cost $c$ to provide her opponent with a benefit $b$. In general, selection can favor cooperation in structured populations provided the benefit-to-cost ratio $b/c$ is sufficiently large \cite{2006-Ohtsuki-p502-505,2017-Allen-p227-230}. Here, we analyze how the critical benefit-to-cost ratio required to support cooperation depends upon directionality in the structure of social interactions.

~\\
We describe population structure by a directed network with $N$ nodes, labelled by $\mathcal{N}=\{1,2,\cdots,N\}$, and edges $w_{ij}$, where $w_{ij}=1$ means an edge from source node $i$ to target node $j$, and $w_{ij}=0$ means no such an edge. 
Note that the values $w_{ij}$ and $w_{ji}$ are not necessary the same.
An edge is bidirected (or undirected) if $w_{ij}=w_{ji}=1$, which is equivalent to having two directed edges in opposite directions. If all edges are bidirected we say the network is undirected.

~\\
Each node is occupied by an individual whose strategy is either cooperate or defect.
Let $s_i$ denote player $i$'s strategy ($s_i=1$ means $C$ and $s_i=0$ meand $D$).
In each generation, a donation game is played along each edge. 
If the player at a source node cooperates, she pays cost $c$ to bring the player at the target node benefit $b$; while a defector at the source node pays no cost and provides no benefit.
If a bidirected edge connects nodes $i$ and $j$ then the donation game is played twice along the edge, with each node assuming the role of potential donor and potential recipient.
After games have been played along all edges, each player accumulates payoffs summed across all interactions, so that the total payoff to $i$ is given by 
\begin{equation}
\pi_i=-c\sum_{j\in\mathcal{N}}w_{ij}s_i+b\sum_{j\in\mathcal{N}}w_{ji}s_j.
\end{equation}
The payoff is then transformed to a reproductive rate $F_i=1+\delta \pi_i$, where $\delta$ denotes the selection intensity \cite{2004-Nowak-p646-650}.
We assume weak selection, $0<\delta\ll 1$, meaning that payoff differences have small effects on the evolutionary dynamics \cite{2004-Nowak-p646-650,2010-Wu-p46106-46106}.. This assumption has a long history in population genetics \cite{Kimura1968} and evolutionary biology \cite{Akashi1995}, and it has also been used to formulate predictions for   behavioral experiments with human subjects \cite{2010-Traulsen-p2962-2966}.

~\\
After receiving payoffs based on their current strategies, individuals have the opportunity to update their strategies by payoff-biased imitation of other players. In general, the network along which strategies spread (or disperse) may not be the same as the network of pairwise game interactions. In particular, at the end of each generation, a random player $i$ is selected uniformally at random to die (or, equivalently, selected to update her strategy), and she imitates the strategy of one of her neighbors in the directed dispersal network, selected proportional to their reproductive rate.

~\\
Although our analysis applies to an arbitrary strategy dispersal network (see Supporting Information), we focus on two specific cases of dispersal networks, termed `downstream' and `upstream' dispersal. The dispersal edges in both the upstream and downstream cases are the same, and they agree with the edges in the pairwise game interaction network. However, in the `downstream' case, the edge directions in the dispersal network are identical to those of the interaction network; and in the `upstream' case the edge directions for strategy dispersal are reversed relative to the directions of the interaction network (see Figure~\ref{fig:1}).
And so, in the downstream case, player $j$ successfully disperses her strategy to $i$ with probability 
\begin{align} \label{ein_jm}
e_{j\rightarrow i} = 
\frac{1}{N}\frac{w_{ji}F_{j}}{\sum_{\ell \in \mathcal{N}} w_{\ell i}F_{\ell}} . 
\end{align}
Whereas in the upstream case, player $j$ successfully disperses her strategy to $i$ with probability 
\begin{align} \label{ein_jm}
e_{j\rightarrow i} = 
\frac{1}{N}\frac{w_{ij}F_{j}}{\sum_{\ell \in \mathcal{N}} w_{i\ell}F_{\ell}} . 
\end{align}

~\\
Loosely speaking, in the downstream case a player at node $i$ acts as a potential donor towards player $j$ in the interaction network, and player $j$ may later choose to imitate the behavior of $i$ and act in the same way towards a third party. In the upstream case, by contrast, a player at node $i$ acts as a potential donor to player $j$, and a third party may later choose to imitate the behavior of $i$ and take it against $i$. When all edges are bidirected, the evolutionary process of interaction and strategic imitation are the same in both upstream and downstream cases, and they coincide with classical models in structured populations \cite{2006-Ohtsuki-p502-505,2017-Allen-p227-230,McAvoy2021}.

\section{Results}
We study the evolution of cooperation by quantifying the chance that a single mutant type, introduced at a random node, will eventually spread and overtake the entire population. 
We assume that the population structure for strategy dispersal is strongly connected, meaning that for any pairs of $i,j\in\mathcal{N}$ there is a directed path from $i$ to $j$ in strategy dispersal networks. Letting $\rho_C$ (respectively $\rho_D$) denote the probability that a single mutant cooperator (respectively defector) eventually overtakes the population, we say that selection favors cooperation over defection provided $\rho_C>\rho_D$ \cite{2004-Nowak-p646-650}.

\subsection{Cooperation can evolve on directed networks}
We start by studying whether cooperation can ever be favored on fully directed networks, which contain no bidirectional edges. To consider this problem we first recall the key intuition used in prior studies to explain why bidirectional population structures can favor cooperation. In that setting, once a cooperator disperses her strategy to a neighbor, she will benefit from reciprocity in the next generation, receiving donations from her neighbor. By contrast, if a defector disperses his strategy to a neighbor, he will suffer in the next generation and receive nothing from his neighbor.
Thus, the formation of `cooperator clusters' and `defector clusters' favors the evolution of cooperation in bidirectional networks \cite{1964-Hamilton-p1-16,1964-Hamilton-jtb2,1992-Nowak-p826-829}. 

~\\
This intuition for the effect of population structure on cooperation relies on reciprocity; and so it does not apply in the setting of directed edges, and especially not in the fully directed setting. For example, in the `downstream' case, if a cooperative player $i$ disperses her strategy along edge $w_{ij}$ to player $j$, then player $j$ is unable to reciprocate $i$, because there is no edge from $j$ to $i$. Likewise, a defector who spreads his strategy will not be retaliated against by defection. And so, lacking the mechanism of network reciprocity, we might expect that directed graphs cannot favor the evolution of cooperation.

\begin{figure*}[!h]
\centering
\includegraphics[width=0.8\textwidth]{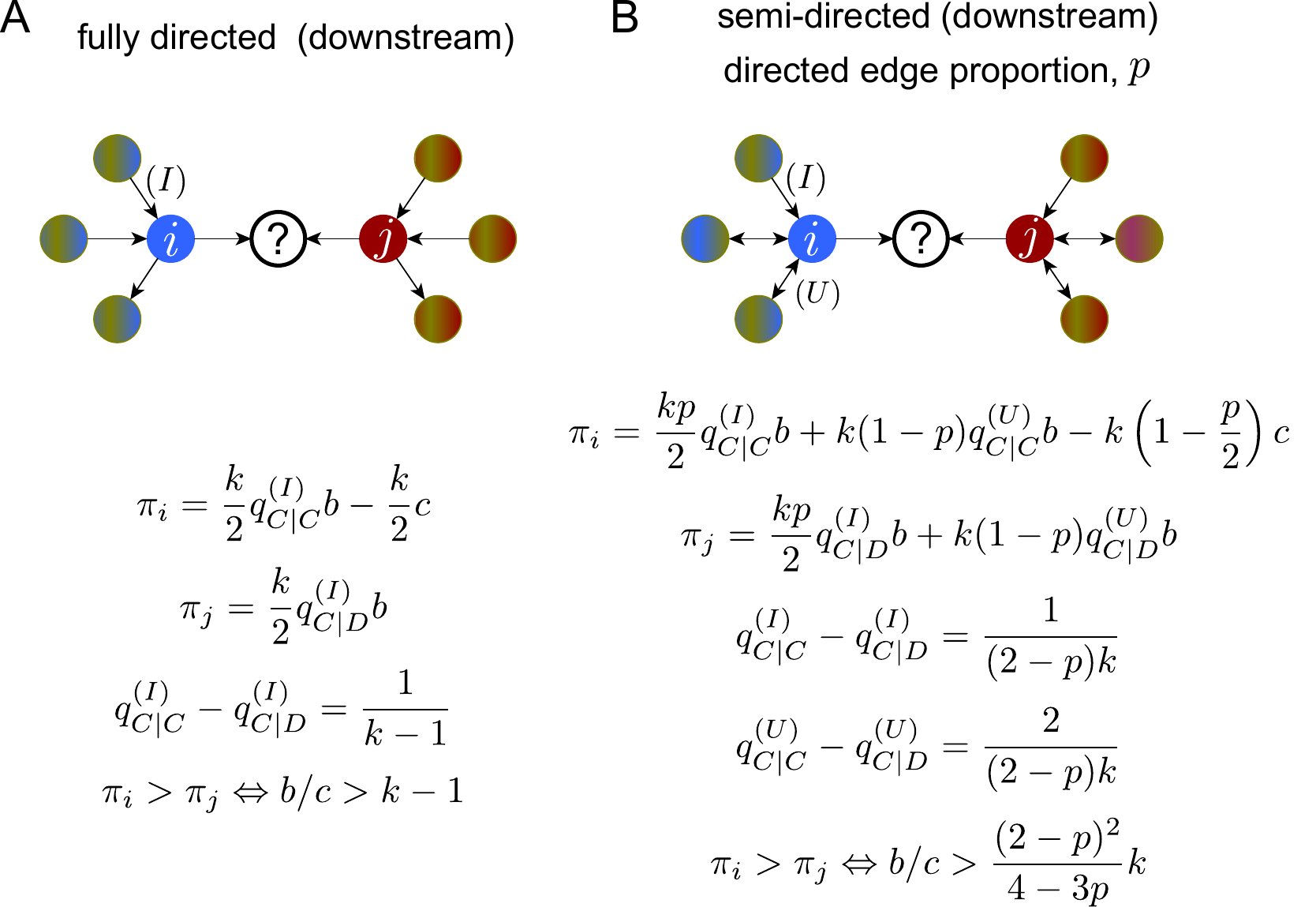}
\caption{\label{fig:2} \textbf{Evolution of cooperation on directed networks.}
When an individual updates her strategy, her neighbors such as cooperator $i$ and defector $j$  compete to disperse their strategy. Cooperation is favored to spread only if the payoff to $i$ exceeds the payoff to $j$, $\pi_i>\pi_j$.
We can prove that a cooperator $i$ has a greater chance of having an incoming cooperative neighbor than a defector does: $q_{C|C}^{(I)} > q_{C|D}^{(I)}$ (see Supporting Information). 
(A) In a fully directed network, cooperator $i$ receives benefits from $k/2$ incoming neighbors and pays costs $c$ to each of $k/2$ outgoing neighbors; whereas defector $j$ receives benefits from $k/2$ incoming neighbors, but pays no costs -- which yields expressions for $\pi_i$ and $\pi_j$. Selection then favors cooperation provided $b/c>k-1$.
(B) In a semi-directed networks with a proportion $p$ of directed edges, $i$ and $j$ each receives potential benefits from $kp/2$ incoming neighbors connected by uni-directed edges ($I$) and from $k(1-p)$ neighbors connected by bi-directed  edges ($U$); and cooperator $i$ pays costs to $k(1-p/2)$ neighbors. Selection favors cooperation provided $b/c>(2-p)^2k/(4-3p)$.
The figure illustrates payoff expectations in the downstream case, for large $k$.
}
\end{figure*}

~\\
Despite the simple intuition above, it turns out that fully directed graphs can indeed favor cooperation. We can prove that in a fully directed random regular network, where each node has $k/2$ incoming edges and $k/2$ outgoing edges, selection favors the evolution of cooperation if (see Supporting Information)
\begin{equation} \label{downstream_fully}
\frac{b}{c}>k-1
\end{equation}
in the `downstream' case and 
\begin{equation} \label{upstream_fully}
\frac{b}{c}>\frac{k(k-1)}{k-2}
\end{equation}
in the `upstream' case, provided the number of nodes $N$ is sufficiently large.

~\\
We can provide some intuitions for this surprising result by considering an individual's expected payoffs over the long-term evolutionary process. When an individual dies (or, equivalently, is chosen to update her strategy), all her incoming neighbors in the dispersal network compete to reproduce and spread their strategy to the vacancy (see Figure \ref{fig:2}A). Cooperation spreads if an incoming cooperator neighbor, such as node $i$, has a higher expected payoff than an incoming defector neighbor, $j$.
Our analysis reveals that such a cooperator $i$ has $k/(2(k-1))$ more incoming cooperative neighbors than a defector $j$ does, and thus $i$ receives $bk/(2(k-1))$ greater benefit than $j$ (see Supporting Information). At the same time, though, the cooperator $i$ pays a cost $ck/2$ along $k/2$ interactions, whereas the defector $j$ avoids these costs. 
And so a cooperator $i$'s net payoff exceeds $j$'s if the benefit exceeds the cost, i.e.~$bk/(2(k-1))>ck/2$ is satisfied, leading to $b/c>k-1$.
We can analyze the `upstream' case in an analogous way. 
The slight difference is that in the `upstream' case, the cooperator $i$ has $(k-2)/(2(k-1))$ more cooperative neighbors than a defector $j$, which is less than in the `downstream' case, and so the evolution of cooperation requires a slightly higher benefit-to-cost ratio.

\subsection{Cooperation thrives when directionality is intermediate}
\begin{figure*}[!h]
\centering
\includegraphics[width=1\textwidth]{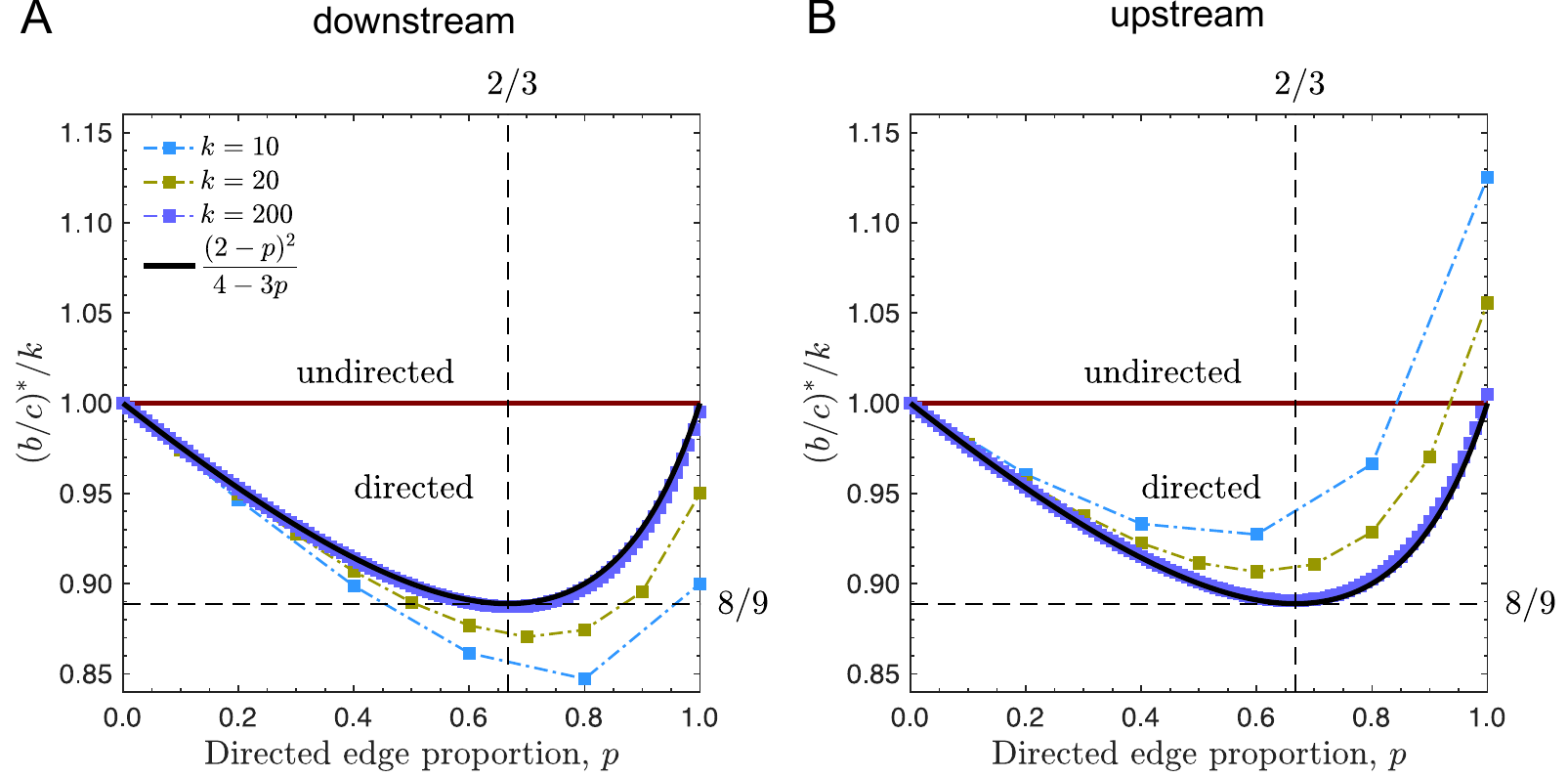}
\caption{\label{fig:3} \textbf{An intermediate proportion of directed edges is optimal for cooperation.} 
We consider random regular networks with a proportion $p$ of directed edges. Each each node has $kp/2$ incoming directed edges, $kp/2$ outgoing directed edges, and $k(1-p)$ undirected edges. The figures display the critical benefit-to-cost ratio $(b/c)^*$ required to favor cooperation, as a function of the edge proportion $p$ and scaled by node degree $k$.  For sufficiently large node degree, cooperation is easiest to evolve (smallest critical ratio) for $p=2/3$, under both downstream (A) and upstream (B) behavior dispersal. The red line indicates the critical ratio for a purely undirected network \cite{2006-Ohtsuki-p502-505}, $(b/c)^*=k$.
}
\end{figure*}

Next we consider the prospects for cooperation on semi-directed networks, consisting of both directed and undirected edges.
Let $p$ denote the proportion of edges that are uni-directional.
In a semi-directed random regular network with  degree $k$ each node has $kp/2$ incoming edges, $kp/2$ outgoing edges, and $k(1-p)$ undirected edges.
For the sufficiently large network size $N$ and large node degree $k$, for both downstream and upstream dispersal, we can prove that selection favors cooperation over defection provided (see Supporting Information)
\begin{equation}
\frac{b}{c}>\frac{(2-p)^2}{4-3p}k.
\end{equation}
When $p=0$ (an undirected network), we recover the classical condition $b/c>k$ \cite{2006-Ohtsuki-p502-505}.
When $p=1$ (a fully directed network), we have $b/c>k$, which approximates Eqs.~(\ref{downstream_fully}) and (\ref{upstream_fully}) for large $k$.
In general, though, an intermediate proportion of directed edges, namely $p=2/3$, minimizes the benefit-to-cost ratio required for selection to favor cooperation 
(see Figure \ref{fig:3}). And so not only is the evolution of cooperation possible on $k$-regular networks whose edges are all directed, but in fact cooperation is made easier when an intermediate portion of edges are directed.

~\\
Figure \ref{fig:2}B provides the key intuition to explain why an intermediate proportion of directed edges is most beneficial for the evolution of cooperation.
We focus on the downstream case, as the upstream case can be analyzed analogously. 
In a semi-directed network, players receive benefits from not only the directed incoming neighbors (marked by $I$), but also from neighbors connected by undirected edges (marked by $U$).
Our analysis shows 
(\romannumeral1) the chance that a cooperator's incoming neighbor is also a cooperator exceeds the chance that a defector's incoming neighbor is a cooperator, i.e. the difference $p_{C|C}^{(I)}-p_{C|D}^{(I)}$ exceeds zero and is monotonically increasing with the proportion of directed edges, $p$;
(\romannumeral2) this probability difference is twice as large for undirected edges than it is for directed incoming edges, i.e. $p_{C|C}^{(U)}-p_{C|D}^{(U)}=2(p_{C|C}^{(I)}-p_{C|D}^{(I)})$. 
Starting from a purely undirected graph ($p=0$), converting some edges to directed ($p>0$)
increases $p_{C|C}^{(U)}-p_{C|D}^{(U)}$ thereby increasing the benefit that a cooperator obtains from undirected neighbors. But as the proportion of directed edges increases yet further, the total number of undirected edges decreases and this reduces the net benefit that the cooperator receives from cooperative neighbors. As a result of these two opposing phenomena, an intermediate proportion of directed edges is optimal: $p>0$ increases the chance that a cooperator's neighbors are cooperators, whereas $p<1$ maintains a sufficient number of undirected neighbors.

~\\
Our analysis of $k$-regular graphs shows that directed interactions can stimulate cooperation, but these results do not account for population heterogeneity, which is common in real-world social networks. To address this, we proceed to investigate four classes of heterogeneous networks: random regular networks in which nodes can have different numbers of incoming,  outgoing, and undirected edges; random networks; small-world networks; and scale-free networks (see Figure S\ref{sfig:2} in Supporting Information).
For each class, we generate undirected networks of various average node degree $\bar{k}$, and then we randomly convert a proportion $p$ of edges to be unidirectional.
For all four classes of heterogeneous networks, we find that
an intermediate proportion of directed edges is most beneficial to cooperation, for both downstream and upstream directions of strategy spread (see Supporting Information for the calculation of benefit-to-cost ratios in any directed network and see Ref.~\cite{McAvoy2021} for details).
For example, for $p=0$ (undirected network), if cooperation can evolve for some benefit-to-cost ratio, increasing $p$ always decreases the ratio required for the evolution of cooperation.
Some undirected networks disfavor cooperation regardless of the benefit-to-cost ratio, or they may even favor the evolution of spite, a kind of antisocial behavior in which an individual pays a cost to hurt others. Even on such undirected networks that disallow cooperation whatsoever, we find that converting some undirected edges to directional can rescue cooperation. Overall, for all four classes of heterogeneous networks, the optimal proportion of directed edges that facilitates cooperation is close to $2/3$, as in the case of random regular networks.

\subsection{Edge orientation matters}
\begin{figure*}[!h]
\centering
\includegraphics[width=1\textwidth]{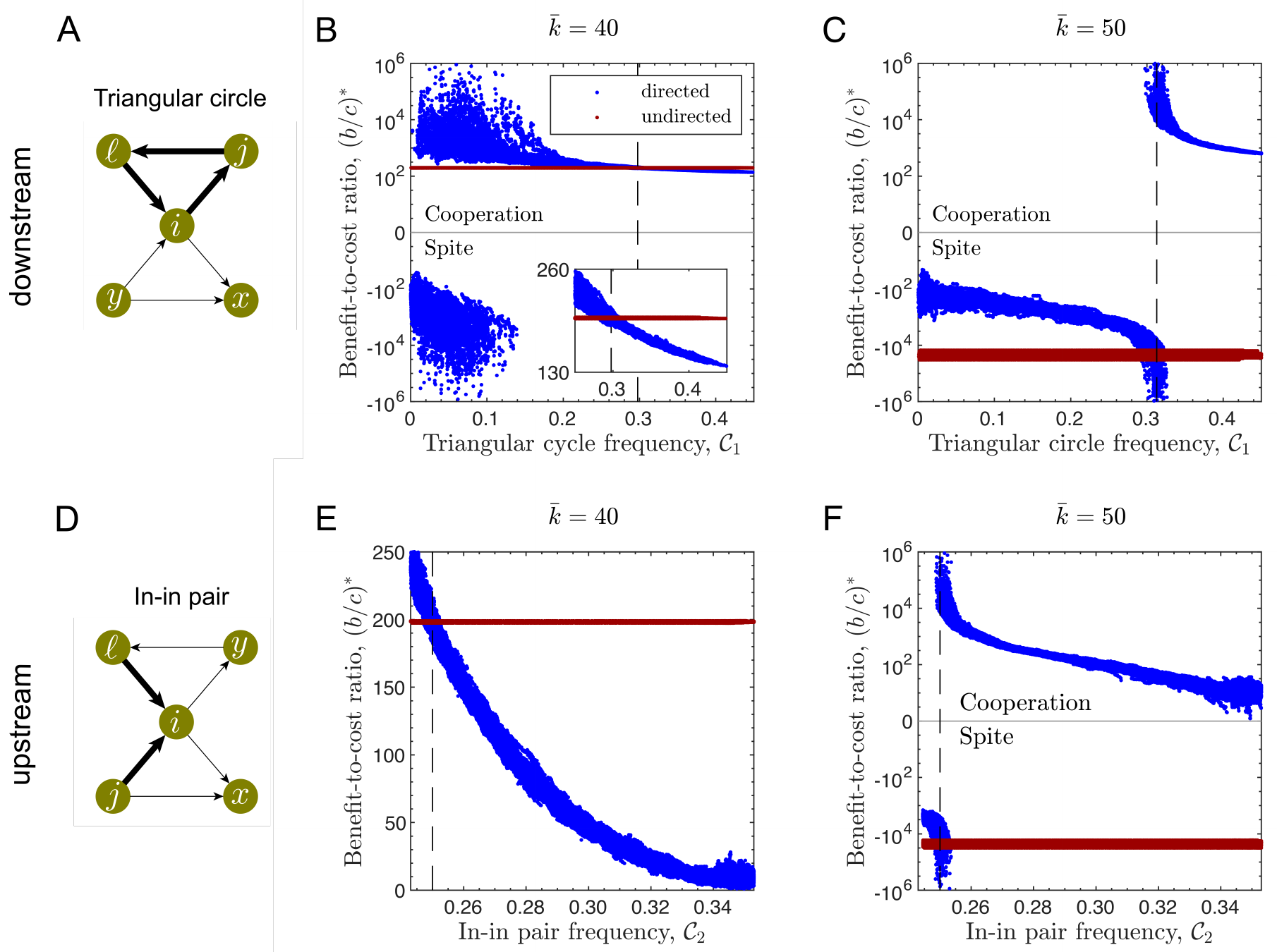}
\caption{\label{fig:4} \textbf{Network motifs that facilitate cooperation.}
We study two network motifs: the triangular cycle for evolutionary dynamics with downstream behavioral dispersal (A), and the in-in pair for upstream dispersal (D). We plot the critical benefit-to-cost ratio required to favor the evolution of cooperation as a function of the frequency of these motifs, $\mathcal{C}_1$ (BC) and 
$\mathcal{C}_2$ (EF). In each case, we start with an undirected Watts-Strogatz small-world network \cite{1998-Watts-p440-442} generated with rewiring probability $0.1$, and we plot the critical ratio in red. We convert all edges to uni-directional, assigning random orientations, and plot the resulting frequency of motifs (vertical lines). By systematically adjusting edge orientations we can either increase or decrease the frequency of motifs, $\mathcal{C}_1$ and 
$\mathcal{C}_2$. A high frequency of triangular cycles facilitates cooperation in the downstream case; and a high frequency of in-in pairs facilitates cooperation in the upstream case. In some cases, when motif frequencies are low or edges are bi-directional, the network favors the evolution of spite, as indicated by a negative value of $(b/c)^*$; adjusting edge directions to increase motif frequency can rescue cooperation. Figures show results for 100 sampled Watts-Strogatz networks, each with $N=100$ nodes. 
For each network, we obtain $10000$ motif frequencies by adjusting edge orientations.
}
\end{figure*}

In the analyses presented so far, when generating a directed network from an undirected network we converted bi-directional edges to directional edges, choosing the orientation of each edge at random.
In this section we explore the strategic assignment of edge directions, to understand how this feature of network topology influences the evolution of cooperation. In particular, we will identify two key network motifs  --- one relevant in the upstream case, and one in the downstream case -- that depend on edge orientation and strongly influence the fate of cooperation in directed networks.
 
~\\
We focus our analysis on fully directed networks, paying attention now to the orientation of each edge.
Let $k_i$ denote node $i$'s degree, including $k_i^{(I)}$ incoming edges and $k_i^{(O)}$ outgoing edges, i.e. $k_i=k_i^{(I)}+k_i^{(O)}$.
For the downstream case, where strategies spread in the same direction as interactions occur, we consider the motif of ``triangular cycles", such as $i\rightarrow j\rightarrow \ell\rightarrow i$ (see Figure \ref{fig:4}A). For node $i$, the number of such triangular cycles is $\sum_{j,\ell}w_{ij}w_{j\ell}w_{\ell i}$.
We normalize the number of such triangles through node $i$ by its maximal value, given the node's in- and out-degree, and define the following quantity
\begin{equation}
\mathcal{C}_1=\frac{\sum_{i,j,\ell} w_{ij}w_{j\ell}w_{\ell i}}{\sum_{i} k_i^{(I)}k_i^{(O)}}
\end{equation}
to measure the (normalized) frequency of triangular cycles in the directed network. A large value of $\mathcal{C}_1$ means there is a large frequency of triangular cycles, given the incoming and outgoing node degrees.

~\\
In the upstream case, where strategies spread in the opposite direction as interactions occur, we consider the motif of ``in-in pairs", such as $j\rightarrow i$ and $\ell \rightarrow i$ for node $i$ in Figure \ref{fig:4}D.
For node $i$, the number of in-in pairs is $k_i^{(I)}\left(k_i^{(I)}-1\right)$. We normalize the number of such in-in pairs by the total number of edge pairs for node $i$ and define the quantity 
\begin{equation}
\mathcal{C}_2=\frac{\sum_i k_i^{(I)}\left(k_i^{(I)}-1\right)}{\sum_i k_i\left(k_i-1\right)}
\end{equation}
to measure the global frequency of in-in pairs in the directed network.
A large value of $\mathcal{C}_2$ means there is a large frequency of in-in pairs, given the incoming and outgoing node degrees.

~\\
For  downstream dispersal, edge orientations that produce a large proportion of triangular cycles are beneficial to the evolution of cooperation (Figure \ref{fig:4}BC). Whereas a directed network with random orientations may require a very high benefit-to-cost ratio for cooperation, adjusting edge directions (only reversing orientations of existing directed edges) to increase the frequency of triangular cycles ($\mathcal{C}_1$) can markedly decrease the benefits required for cooperation (see Figure \ref{fig:4}B).
Furthermore, even when a directed network with random orientations disfavors cooperation altogether, for any benefit-to-cost ratio, adjusting edge orientations to increase triangular cycles can rescue cooperation (Figure \ref{fig:4}C). 

~\\
In-in pairs have analogous, beneficial effects on cooperation in the case of upstream strategy dispersal (Figure \ref{fig:4}EF). Moreover, the effects of triangular cycles and in-in pairs, in the downstream and upstream contexts respectively, persist across a large sample of regular, random, and semi-directed networks, even when connections are sparse (see Figures \ref{sfig:4}-\ref{sfig:6} in Supporting Information).
The only counterexample occurs when heterogeneity in node degree is extremely large (such as a heavy-tailed degree distribution), which mitigates the cooperation-promoting effects of triangular cycles in the downstream setting.

\begin{figure*}[!h]
\centering
\includegraphics[width=0.95\textwidth]{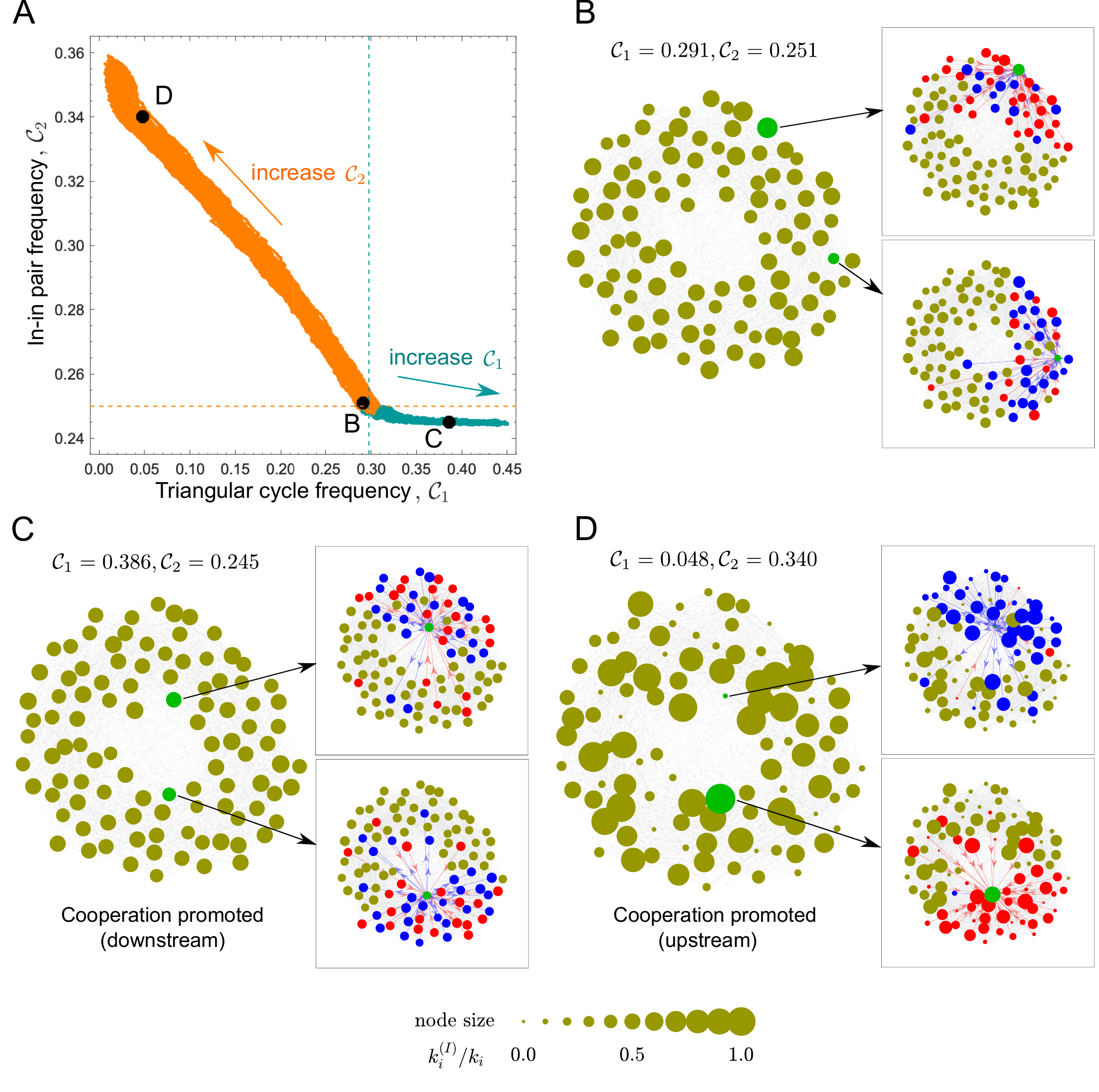}
\caption{\label{fig:5}  \textbf{Motif frequency and node degrees}.
We generated $100$ Watts-Strogatz small-world networks \cite{1998-Watts-p440-442} with size $N=100$, average degree $\bar{k}=40$, and 
rewiring probability $0.1$. (A) After making all edges uni-directional and assigning random orientations, we measured the frequency of triangular cycles and in-in pairs (vertical and horizontal lines). 
We then modified edge directions either to increase $\mathcal{C}_1$ (teal)  or to increase $\mathcal{C}_2$ (orange), recording both $\mathcal{C}_1$ and $\mathcal{C}_2$, which are anti-correlated. Panels B, C, and D illustrate three example networks with intermediate, high, and low frequency of triangular cycles, as indicated on the network of panel A.
For each example network, the size of each node $i$ is proportional to its in-degree relative to its total degree,  $k_i^{(I)}/k_i$.
In each network, the connections of two representative nodes (green circle) are highlighted, with red circles denoting incoming neighbors and blue circles outgoing neighbors. A large frequency of triangular cycles is associated with homogeneity of in-degree/out-degree (C), and this tends to promote cooperation under downstream behavioral dispersal (see Figure \ref{fig:4}). Whereas a large frequency of in-in pairs is associated with heterogeneity of in-degrees/out-degrees (D), and this tends to promote cooperation in the case of  upstream dispersal (see Figure \ref{fig:4}).
}
\end{figure*}

~\\
There is a simple intuition for why triangular cycles promote cooperation in the downstream setting, and in-in pairs promote cooperation in the upstream setting. Although immediate reciprocity is impossible on a fully directed network in the downstream case, the triangular cycle $i\rightarrow j\rightarrow \ell\rightarrow i$ allows cooperator $i$ to  receive reciprocity via two-step strategy dispersal: cooperation spreads from $i$ to $j$ and from $j$ to $\ell$, and then $i$ is reciprocated by $\ell$.
This effect is second-best to pairwise reciprocity, and orienting edges to produce triangular cycles proves to be efficient for promoting cooperation. In the upstream case, by contrast, a cooperator at node $i$ can be reciprocated immediately once she successfully disperses cooperation to an incoming neighbor, such as $j\rightarrow i$. Therefore, if a focal cooperator has more incoming neighbors, the benefits of reciprocity are are strengthened. Moreover, reciprocity arising from multiple incoming neighbors apparently has synergistic effects on cooperation -- because the frequency of in-in pairs is strongly correlated with the emergence of cooperation, whereas the number incoming edges alone does not vary with edge orientations.

~\\
The two motifs, triangular cycles and in-in pairs, are often mutually exclusive.
Increasing the frequency of triangular cycles, $\mathcal{C}_1$, by re-orienting edge directions tends to decrease the frequency of in-in pairs,  $\mathcal{C}_2$, and vice versa (see Figure \ref{fig:5}A). Figure \ref{fig:5}BCD illustrates three directed networks with various frequencies of these motifs, while also showing each node's in-degree relative to its total degree $k_i^{(I)}/k_i$ (see also Figure \ref{sfig:3} in Supporting Information). As these examples illustrate, when edge orientations are optimized to increase the frequency of triangular cycles (panel C), this has the effect of homogenizing the in-degree and out-degree across the network. Whereas optimizing the frequency of in-in pairs (panel D) leads to extreme heterogeneity in the in-degree/total-degree ratios across nodes -- so that a few nodes serve primarily as sources, and a few nodes primarily as sinks. These qualitative observations provide some intuitions for what features of degree heterogeneity across a directed network are likely to stimulate cooperation, for either the downstream case of strategy imitation (homogeneous in-degrees) or upstream case (heterogeneous in-degrees).


\begin{table}
\begin{adjustwidth}{0in}{0in}
\centering
\footnotesize
\begin{tabular}{m{2.0cm}<{\centering}| m{1.6cm}<{\centering} m{1.6cm}<{\centering} m{1.2cm}<{\centering} m{1.2cm}<{\centering} m{1.2cm}<{\centering} m{1.2cm}<{\centering} m{1.2cm}<{\centering} m{1.2cm}<{\centering}}
\toprule
\bf{Name} & \bf{Nodes} & \bf{Social ties} & \bm{$p$} & $\bm{\left(b/c\right)_{\text{un}}^{\ast}}$ & $\bm{\left(b/c\right)^{\ast}_{\text{emp,d}}}$ & $\bm{\left(b/c\right)^{\ast}_{\text{emp,u}}}$ & $\bm{\left(b/c\right)^{\ast}_{\text{rand,d}}}$ & $\bm{\left(b/c\right)^{\ast}_{\text{rand,u}}}$  \\
\midrule
Families in Costa Rica \cite{nooy-2018} & families (40) & visiting ties & 0.45 & 3.95 & 3.88 &  3.20 & 3.84 & 3.53\\
\midrule
Australian National University campus \cite{Freeman1998} & students (16) & friendship ratings & 0.54 & 21.49 & 19.82 & 15.86 & 
19.23 & 18.31\\
\midrule
Physicians in Illinois \cite{Coleman1957} & physicians (95) & trust, advice  & 0.80 & 10.41 & 8.32 & 5.07 &
9.75 & 8.38\\
\midrule
Twitter users \cite{DeChoudhury2010} & users (1726) & following relationship  & 0.79 & 8.83 & 6.47 & 3.06 &
9.37 & 6.79\\
\bottomrule
\end{tabular}
\end{adjustwidth}
\caption{\textbf{Evolution of cooperation in four empirical directed networks.} 
In the family visitation network of San Juan Sur, Costa Rica, a directed edge means that one family (source node) visits another family (target node) \cite{nooy-2018}.
In the friendship rating network of residents on the Australian National University campus, a directed edge means that the source node rates the target node as a friend \cite{Freeman1998}.
In the physician trust network in Illinois, a directed edge means that the physician at the source node trusts or asks for professional advice from the target node \cite{Coleman1957}.
In the Twitter follower network (based on a snowball sample crawl across ``quality'' users in 2009), a directed edge means that the source node follows the target node \cite{DeChoudhury2010}.
We extract the largest strongly connected component of each network and treat all edges with weight one. 
The table summarizes the proportion of directed edges ($p$) and the critical benefit-to-cost ratio required to favor cooperation for the empirical networks assuming downstream strategy dispersal ($(b/c)^*_{\text{emp,d}}$) or upstream dispersal ($(b/c)^*_{\text{emp,u}}$). For comparison, the table also presents the critical ratio after removing all directionality information and treating all edges as bi-directional ($(b/c)^*_{\text{un}}$), or after randomizing the orientation of directional edges ($(b/c)^*_{\text{rand,d}}$ and $(b/c)^*_{\text{rand,u}}$, averaged over $100$ random assignments of orientations). 
\label{tab:1}}
\end{table}
~\\

\subsection{Cooperation on empirical social networks}

Finally, we investigated strategic evolution on four empirical social networks that feature directed edges (Table \ref{tab:1}). These networks were assembled from empirical data on physical visitations among families in a Costa Rican town \cite{nooy-2018}, surveys of friendship ratings on a University campus \cite{Freeman1998}, the referral network among a sample of physicians in Illinois \cite{Coleman1957}, and the follower-network among a sample of Twitter users \cite{DeChoudhury2010}. The proportion of unidirectional edges ranges from $p=45\%$ to $p=80\%$ in these empirical examples. 

~\\
In these four empirical networks, cooperation is easier to evolve under our model when accounting for the empirical directionality of edges, under either upstream or downstream strategic spread, as compared to a model that ignores the empirical directionality and treats all edges as bidirectional. In particular, respecting the directionality measured in these empirical settings can substantially reduce the benefit-to-cost ratio required for cooperation, by two-fold or more (Table \ref{tab:1}). Moreover, the actual orientation of directional edges on most empirical networks also tends to favors cooperation, compared to randomly re-orientating edges. As these results show, the benefits of directional social interactions for cooperation are not merely theoretical predictions for stylized abstract networks, but they also arise from the directionality measured in empirical networks of real-world social interactions.

\section{Discussion}
Population structure has long been recognized as a catalyst for cooperation that cannot otherwise spread in a well-mixed society. And yet most theoretical analysis of this effect has assumed bi-directional social interactions, even though real-world interactions are often uni-directional. Directionality arises in real-life settings as the result of organizational hierarchy, social stratification, popularity effects, as well as endogenous mechanisms of network growth in  online social networks. But the impact of directed interactions for cooperation have not been thoroughly studied, by either theoretical analysis or empirical experimentation. One reason why directionality may have been neglected is that, a priori, directional interactions would seem to only impede cooperation, because they remove the possibility of pairwise reciprocity \cite{1964-Hamilton-p1-16,1964-Hamilton-jtb2,1992-Nowak-p826-829}.


~\\
Our results contravene the simple intuition that directionality should impede cooperation. Even though it disrupts reciprocity, we have proven analytically that cooperation can be favored to evolve in populations with purely directional social interactions and directional spread of behavior. This analysis is based on the same theoretical framework that has been widely applied to study the emergence of cooperation in undirected social structures \cite{2006-Ohtsuki-p502-505,2007-Ohtsuki-p108106-108106}. Our results rest on the simple intuition that cooperators tend to have more cooperative neighbors than defectors do -- and that, in fact, this effect is maximized when an intermediate proportion of edges in the population are uni-directional.

~\\
Although we have focused on the donation game as a well established model for a social dilemma \cite{sigmund-2010}, our method of analysis provides general conditions for the evolution of cooperation for arbitrary asymmetric pairwise games (see Supporting Information).

~\\
A recent, systematic study of undirected networks found that cooperation will be disfavored, regardless of the benefit-to-cost ratio, for roughly one-third of network structures \cite{2017-Allen-p227-230}. Many of these structures even favor the evolution of spite \cite{2017-Allen-p227-230}. And yet we have found that, even for such networks, conversion of some bidirectional edges to uni-directional can rescue cooperation. An important implication is that directionality provides an alternative method of modifying population structure to promote cooperation --- besides severing old ties and building new ones, as has been explored in the undirected setting \cite{2017-Allen-p227-230,Fotouhi2018a,2020-qi-arxiv}. In particular, we have identified two specific structural motifs, triangular cycles for downstream behavior dispersal and in-in pairs for upstream dispersal, that efficiently guide structural modifications that favor the spread of cooperative behavior.

~\\
Several prior studies have considered the role of directionality in population structure on the spread of mutant types \cite{Masuda2009,Pavlogiannis2018,2019-Moeller-p137-137,Tkadlec2020}.
But in those studies the fitness advantage of the mutant is fixed, independent of type frequency. By contrast, our study considers frequency-dependent fitness effects that therefore describe game-theoretic interactions: an individual's payoff depends on both her type and her neighbors' types. The question of directionality in models of social behavior has been analyzed in at least one prior paper \cite{2005-Lieberman-p312-316}, but only in the case of two specific networks.  Lieberman $et. al$ analyzed a directed circle network, where each node has one incoming neighbor and one outgoing neighbor, and a super-star network \cite{2005-Lieberman-p312-316}.
Whereas that work shows that directionality can influence the spread of cooperation, in those two specific cases the effect of directionality is  to repress cooperation.
Our analysis applies to arbitrary network structures and we find that, in general, directionality tends to favor cooperation. Other studies have considered distinct networks of social interactions versus behavioral dispersal, limited to purely bi-directional links \cite{2007-Ohtsuki-p108106-108106,2019-Qi-p20190041-20190041}.

~\\
Our results reveal the importance of directed interactions for the long-term prospects of cooperative behavior in a population. 
Our analysis also highlights an interesting relation between the direction of social interactions versus the direction of behavioral spread (or dispersal). We have focused on two opposite extremes: downstream dispersal, in which behavior spreads in the same direction as directed social interactions, and upstream dispersal, in which behavior spreads in the opposite direction. In the downstream setting, the directional motif $i\rightarrow j$ and $j\rightarrow \ell$ indicates that a cooperator $i$ donates to $j$, and $j$ might then imitate $i$'s cooperative behavior and make a donation to $\ell$. This form of dispersal is roughly analogous to generalized reciprocity --- a player who received help from another player feels motivated to help the third player in turn \cite{Hamilton2005}. In the upstream setting, by contrast, the directional motif $i\rightarrow j$ and $\ell \rightarrow i$ indicates that cooperator $i$ donates to $j$, and $\ell$ might then imitates $i$'s cooperative behavior and make a donation to $i$. This form of behavioral dispersal is roughly analogous to indirect reciprocity --- the player who helped another player receives a benefit from a third party \cite{2005-Nowak-p1291-1298}.
Despite these rough similarities and their beneficial effects for cooperation, the form of reciprocity arising from directed dispersal of behavior differs from generalized reciprocity and from indirect reciprocity: the dispersal mechanism is payoff-dependent while the later mechanisms are action-dependent. A synthetic understanding of reciprocity based on conditional behavior versus reciprocity achieved by payoff-biased imitation remains an important topic for future research.



\clearpage
\begin{center} \bfseries \Large \singlespacing
Evolution of cooperation with asymmetric social interactions\\
		\textit{Supporting Information} \\ \bigskip 
\end{center}
\setcounter{equation}{0}
\setcounter{figure}{0}
\setcounter{section}{0}
\setcounter{table}{0}
\renewcommand{\theequation}{SI.\arabic{equation}}
\renewcommand{\thefigure}{SI.\arabic{figure}}
\renewcommand{\thesection}{SI.\arabic{section}}
\renewcommand{\thesubsection}{SI.\arabic{section}.\arabic{subsection}}
\renewcommand{\thetable}{SI.\arabic{table}}

The population structure is described by a directed network with $N$ nodes (labelled by $\mathcal{N}=\{1,2,\cdots,N\}$) and directed edges $w_{ij}$, where $w_{ij}>0$ means a directed edge from source node $i$ to target node $j$.
To ensure absorption of one type or another, the directed network is required to be strongly-connected.
That is, for all pairs of $i,j\in\mathcal{N}$, there is a path from $i$ to $j$. The payoff structure for the interaction in directed edge $w_{ij}$ is 
\begin{equation}
\bordermatrix{
  & \textit{A} & \textit{B} \cr
\textit{A} & a_1,a_2 & b_1,b_2 \cr
\textit{B} & c_1,c_2 & d_1,d_2 \cr
}.
\end{equation}
The entry $(X,Y)$ in the payoff matrix means that when the player at the source node (i.e. $i$) uses the strategy in the row and the player in the target node (i.e. $j$) uses the strategy in the column, the former obtains payoff $X$ and the later obtains payoff $Y$.
Let $s_i$ denote player $i$'s strategy ($s_i=1$ means $A$-strategy and $s_i=0$ means $B$-strategy).
The accumulated payoff for player $i$ is 
\begin{equation}
\begin{split}
\pi_i = &\sum_{j}w_{ij}\left[a_1s_is_j+b_1s_i(1-s_j)+c_1(1-s_i)s_j+d_1(1-s_i)(1-s_j)\right] \\
&+\sum_{j}w_{ji}\left[a_2s_is_j+b_2(1-s_i)s_j+c_2s_i(1-s_j)+d_2(1-s_i)(1-s_j)\right],
\end{split}
\end{equation} 
which is then transformed to a reproductive rate by $F_i=1+\delta \pi_i$.
At the end of each generation, a random player $i$ is selected to ``die" uniformly. 
Then all players that occupy source nodes of $i$'s incoming edges complete to reproduce an offspring and replace the vacancy at $i$, with probability proportional to their reproductive rate.
Therefore, player $j$ successfully sends an ``offspring" to $i$ with probability 
\begin{align} \label{ein_jm}
e_{j\rightarrow i} = 
\frac{1}{N}\frac{w_{ji}F_{j}}{\sum_{\ell \in \mathcal{N}} w_{\ell i}F_{\ell}} . 
\end{align}
The process described above corresponds to the `downstream' dispersal in the main text.
The equations for `upstream' dispersal can be easily modified from the `downstream' case.

\section{Methods}
We begin with a semi-directed and unweighted random regular network, in which each node has $k^{(I)}$ incoming edges, $k^{(O)}$ outgoing edges, and $k^{(U)}$ undirected (bi-directed) edges.
We have $k^{(I)}=k^{(O)}$ and node degree $k=k^{(I)}+k^{(O)}+k^{(U)}$.
Let $p_A$ (resp. $p_B$) denote the frequency of $A$-players (resp. $B$-players).
Each player therefore thus has three types of neighbors, namely incoming neighbors ($I$, source nodes of incoming edges), outgoing neighbors ($O$, target nodes of outgoing edges), and undirected/bidirected neighbors ($U$, neighboring nodes of undirected/bidirected edges).
Let $p_{XY}^{(I)}$ denote the frequency of incoming edges $X\leftarrow Y$ (taking the first subscript as the focal player);
$p_{XY}^{(O)}$ the frequency of outgoing edges $X\rightarrow Y$;
$p_{XY}^{(U)}$ the frequency of bidirected edges $X\leftrightarrow Y$.
Let $q_{Y|X}^{(I)}$ denote the probability that given the focal player is an $X$-player, the incoming neighbor is a $Y$-player;
$q_{Y|X}^{(O)}$ the probability that given the focal player is an $X$-player, the outgoing neighbor is a $Y$-player;
$q_{Y|X}^{(U)}$ the probability that given the focal player is an $X$-player, the bidirected neighbor is a $Y$-player.
We have the following identities
\begin{equation}
\begin{split}
p_A+p_B & = 1, \\
p_{AB}^{(Z)} & = p_{BA}^{(Z)},  \\
q_{X|Y}^{(Z)} & = \frac{p_{XY}^{(Z)}}{p_Y},  \\
q_{A|Y}^{(Z)}+q_{B|Y}^{(Z)} & = 1,  
\end{split}
\end{equation}
where $X,Y \in\{A,B\}$ and $Z\in\{I,O,U\}$.
Furthermore, we have
\begin{align}
p_{AA}^{(I)}=p_Aq_{A|A}^{(I)}=p_{AA}^{(O)}=p_Aq_{A|A}^{(O)},
\end{align}
which gives $q_{A|A}^{(I)}=q_{A|A}^{(O)}$ and
\begin{align}
p_{AB}^{(I)}=p_Aq_{B|A}^{(I)}=p_A\left(1-q_{A|A}^{(I)}\right)=p_A\left(1-q_{A|A}^{(O)}\right)=p_Aq_{B|A}^{(O)}=p_{AB}^{(O)}=p_{BA}^{(O)}.
\end{align}
Overall, the whole system can be described by three variables, i.e. $p_A$, $q_{A|A}^{(I)}$, and $q_{A|A}^{(U)}$.

\subsection{Updating a B-player}

We first investigate the case where a $B$-player is replaced by a neighboring $A$-player.
Let $k_A^{(Z)}$ and $k_B^{(Z)}$ denote the numbers of $A$- and $B$-players among three types of neighbors, $Z\in\{I,O,U\}$.
We have $k_A^{(Z)}+k_B^{(Z)}=k^{(Z)}$ and $\sum_{Z\in\{I,O,U\}}\left(k_A^{(Z)}+k_B^{(Z)}\right)=k$.
Such a neighborhood configuration occurs with probability
\begin{align}
\mathcal{B}\left(k_A^{(I)},k_A^{(O)},k_A^{(U)}\right)
=\prod_{Z\in \{I,O,U\}}{k^{(Z)} \choose k_A^{(Z)}}\left(q_{A|B}^{(Z)}\right)^{k_A^{(Z)}}\left(q_{B|B}^{(Z)}\right)^{k_B^{(Z)}}.
\end{align}
We introduce two quantities 
\begin{equation}
\begin{split}
\pi_A &= \left(k^{(I)}q_{A|A}^{(I)}+k^{(U)}q_{A|A}^{(U)}\right)(a_1+a_2)
+\left(k^{(I)}q_{B|A}^{(I)}+k^{(U)}q_{B|A}^{(U)}\right)(b_1+c_2), \\
\pi_B &= \left(k^{(I)}q_{A|B}^{(I)}+k^{(U)}q_{A|B}^{(U)}\right)(c_1+b_2)
+\left(k^{(I)}q_{B|B}^{(I)}+k^{(U)}q_{B|B}^{(U)}\right)(d_1+d_2). 
\end{split}
\end{equation}
The average fitness of each $A$- and $B$-player neighbor of type $Z$ is 
\begin{align}
&F_{A|B}^{(Z)} = 1+\delta\pi_{A|B}^{(Z)},\\
&F_{B|B}^{(Z)} = 1+\delta\pi_{B|B}^{(Z)},
\end{align}
where
\begin{equation}
\begin{split}
&\pi_{A|B}^{(I)}=\pi_A-q_{A|A}^{(O)}a_1-q_{B|A}^{(O)}b_1+b_1,\\
&\pi_{B|B}^{(I)}=\pi_B-q_{A|B}^{(O)}c_1-q_{B|B}^{(O)}d_1+d_1,\\
&\pi_{A|B}^{(U)}=\pi_A-q_{A|A}^{(U)}(a_1+a_2)-q_{B|A}^{(U)}(b_1+c_2)+b_1+c_2,\\
&\pi_{B|B}^{(U)}=\pi_B-q_{A|B}^{(U)}(c_1+b_2)-q_{B|B}^{(U)}(d_1+d_2)+d_1+d_2.
\end{split}
\end{equation}
Under such a neighborhood configuration, the probability that an $A$-player takes over the empty site is
\begin{align}
\frac{k_A^{(I)}F_{A|B}^{(I)}+k_A^{(U)}F_{A|B}^{(U)}}{k_A^{(I)}F_{A|B}^{(I)}+k_B^{(I)}F_{B|B}^{(I)}+
k_A^{(U)}F_{A|B}^{(U)}+k_B^{(U)}F_{B|B}^{(U)}}.
\end{align}
Therefore, $p_A$ increases by $1/N$ with probability
\begin{align}
\text{Prob}\left(\Delta p_A=\frac{1}{N}\right)=p_B\sum_{k_A^{(I)},k_A^{(O)},k_A^{(U)}}\mathcal{B}\left(k_A^{(I)},k_A^{(O)},k_A^{(U)}\right)
\frac{k_A^{(I)}F_{A|B}^{(I)}+k_A^{(U)}F_{A|B}^{(U)}}{k_A^{(I)}F_{A|B}^{(I)}+k_B^{(I)}F_{B|B}^{(I)}+
k_A^{(U)}F_{A|B}^{(U)}+k_B^{(U)}F_{B|B}^{(U)}}. \nonumber
\end{align}
The number of incoming edge $A\leftarrow A$ increases by $k_A^{(I)}+k_A^{(O)}$.
Therefore $p_{AA}^{(I)}$ increases by $\left(k_A^{(I)}+k_A^{(O)}\right)/\left(k^{(I)}N\right)$ with probability
\begin{align}
\text{Prob}\left(\Delta p_{AA}^{(I)}=\frac{k_A^{(I)}+k_A^{(O)}}{k^{(I)}N}\right)=p_B\mathcal{B}\left(k_A^{(I)},k_A^{(O)},k_A^{(U)}\right)
\frac{k_A^{(I)}F_{A|B}^{(I)}+k_A^{(U)}F_{A|B}^{(U)}}{k_A^{(I)}F_{A|B}^{(I)}+k_B^{(I)}F_{B|B}^{(I)}+
k_A^{(U)}F_{A|B}^{(U)}+k_B^{(U)}F_{B|B}^{(U)}}. \nonumber
\end{align}
The number of bidirected edge $A\leftrightarrow A$ increases by $k_A^{(U)}$.
Therefore $p_{AA}^{(U)}$ increases by $2k_A^{(U)}/\left(k^{(U)}N\right)$ with probability
\begin{align}
\text{Prob}\left(\Delta p_{AA}^{(U)}=\frac{2k_A^{(U)}}{k^{(U)}N}\right)=p_B\mathcal{B}\left(k_A^{(I)},k_A^{(O)},k_A^{(U)}\right)
\frac{k_A^{(I)}F_{A|B}^{(I)}+k_A^{(U)}F_{A|B}^{(U)}}{k_A^{(I)}F_{A|B}^{(I)}+k_B^{(I)}F_{B|B}^{(I)}+
k_A^{(U)}F_{A|B}^{(U)}+k_B^{(U)}F_{B|B}^{(U)}}. \nonumber
\end{align}

\subsection{Updating an A-player}
We then investigate the case where an $A$-player is replaced by a neighboring $B$-player.
Let $k_A^{(Z)}$ and $k_B^{(Z)}$ denote the numbers of $A$- and $B$-players among the three types of neighbors, $Z\in\{I,O,U\}$.
Such a neighborhood configuration occurs with probability
\begin{align}
\mathcal{A}\left(k_A^{(I)},k_A^{(O)},k_A^{(U)}\right)
=\prod_{Z\in \{I,O,U\}}{k^{(Z)} \choose k_A^{(Z)}}\left(q_{A|A}^{(Z)}\right)^{k_A^{(Z)}}\left(q_{B|A}^{(Z)}\right)^{k_B^{(Z)}}.
\end{align}
The average fitness of each $A$- and $B$-player is given by
\begin{align}
&F_{A|A}^{(Z)} = 1+\delta\pi_{A|A}^{(Z)},\\
&F_{B|A}^{(Z)} = 1+\delta\pi_{B|A}^{(Z)},
\end{align}
where
\begin{equation}
\begin{split}
&\pi_{A|A}^{(I)}=\pi_A-q_{A|A}^{(O)}a_1-q_{B|A}^{(O)}b_1+a_1,\\
&\pi_{B|A}^{(I)}=\pi_B-q_{A|B}^{(O)}c_1-q_{B|B}^{(O)}d_1+c_1,\\
&\pi_{A|A}^{(U)}=\pi_A-q_{A|A}^{(U)}(a_1+a_2)-q_{B|A}^{(U)}(b_1+c_2)+a_1+a_2,\\
&\pi_{B|A}^{(U)}=\pi_B-q_{A|B}^{(U)}(c_1+b_2)-q_{B|B}^{(U)}(d_1+d_2)+c_1+b_2.
\end{split}
\end{equation}
Under such a neighborhood configuration, the probability that an $A$-player takes over the empty site is
\begin{align}
\frac{k_B^{(I)}F_{B|A}^{(I)}+k_B^{(U)}F_{B|A}^{(U)}}{k_A^{(I)}F_{A|A}^{(I)}+k_B^{(I)}F_{B|A}^{(I)}+
k_A^{(U)}F_{A|A}^{(U)}+k_B^{(U)}F_{B|A}^{(U)}}.
\end{align}
Therefore, $p_A$ decreases by $1/N$ with probability
\begin{align}
\text{Prob}\left(\Delta p_A=-\frac{1}{N}\right)=p_A\sum_{k_A^{(I)},k_A^{(O)},k_A^{(U)}}\mathcal{A}\left(k_A^{(I)},k_A^{(O)},k_A^{(U)}\right)
\frac{k_B^{(I)}F_{B|A}^{(I)}+k_B^{(U)}F_{B|A}^{(U)}}{k_A^{(I)}F_{A|A}^{(I)}+k_B^{(I)}F_{B|A}^{(I)}+
k_A^{(U)}F_{A|A}^{(U)}+k_B^{(U)}F_{B|A}^{(U)}}. \nonumber
\end{align}
The number of incoming edge $A\leftarrow A$ decreases by $k_A^{(I)}+k_A^{(O)}$.
Therefore $p_{AA}^{(I)}$ decreases by $\left(k_A^{(I)}+k_A^{(O)}\right)/\left(k^{(I)}N\right)$ with probability
\begin{align}
\text{Prob}\left(\Delta p_{AA}^{(I)}=-\frac{k_A^{(I)}+k_A^{(O)}}{k^{(I)}N}\right)=p_A\mathcal{A}\left(k_A^{(I)},k_A^{(O)},k_A^{(U)}\right)
\frac{k_B^{(I)}F_{B|A}^{(I)}+k_B^{(U)}F_{B|A}^{(U)}}{k_A^{(I)}F_{A|A}^{(I)}+k_B^{(I)}F_{B|A}^{(I)}+
k_A^{(U)}F_{A|A}^{(U)}+k_B^{(U)}F_{B|A}^{(U)}}. \nonumber
\end{align}
The number of bidirected edge $A\leftrightarrow A$ decreases by $k_A^{(U)}$.
Therefore $p_{AA}^{(U)}$ decreases by $2k_A^{(U)}/\left(k^{(U)}N\right)$ with probability
\begin{align}
\text{Prob}\left(\Delta p_{AA}^{(U)}=-\frac{2k_A^{(U)}}{k^{(U)}N}\right)=p_A\mathcal{A}\left(k_A^{(I)},k_A^{(O)},k_A^{(U)}\right)
\frac{k_B^{(I)}F_{B|A}^{(I)}+k_B^{(U)}F_{B|A}^{(U)}}{k_A^{(I)}F_{A|A}^{(I)}+k_B^{(I)}F_{B|A}^{(I)}+
k_A^{(U)}F_{A|A}^{(U)}+k_B^{(U)}F_{B|A}^{(U)}}. \nonumber
\end{align}

\subsection{Separation of time scales}
Assuming that one replacement event happens in one unit of time, we have the derivatives of $p_A$, $p_{AA}^{(I)}$, and $p_{AA}^{(U)}$, given by
\begin{equation}\label{EpA}
\begin{split}
\dot{p}_A
=& \frac{1}{N}\cdot\text{Prob}\left(\Delta p_A=\frac{1}{N}\right)+\left(-\frac{1}{N}\right)\cdot\text{Prob}\left(\Delta p_A=-\frac{1}{N}\right)  \\
=& \frac{\delta}{N\left(k^{(I)}+k^{(U)}\right)^2}\left[
p_B\left(\substack{
k^{(I)}(k^{(I)}-1)q_{A|B}^{(I)}q_{B|B}^{(I)}\left(\pi_{A|B}^{(I)}-\pi_{B|B}^{(I)}\right) \\ 
+k^{(I)}k^{(U)}q_{A|B}^{(I)}q_{B|B}^{(U)}\left(\pi_{A|B}^{(I)}-\pi_{B|B}^{(U)}\right) \\
+k^{(I)}k^{(U)}q_{A|B}^{(U)}q_{B|B}^{(I)}\left(\pi_{A|B}^{(U)}-\pi_{B|B}^{(I)}\right) \\
k^{(U)}(k^{(U)}-1)q_{A|B}^{(U)}q_{B|B}^{(U)}\left(\pi_{A|B}^{(U)}-\pi_{B|B}^{(U)}\right)
}\right) 
-p_A\left(\substack{
k^{(I)}(k^{(I)}-1)q_{B|A}^{(I)}q_{A|A}^{(I)}\left(\pi_{B|A}^{(I)}-\pi_{A|A}^{(I)}\right) \\ 
+k^{(I)}k^{(U)}q_{B|A}^{(I)}q_{A|A}^{(U)}\left(\pi_{B|A}^{(I)}-\pi_{A|A}^{(U)}\right) \\
+k^{(I)}k^{(U)}q_{B|A}^{(U)}q_{A|A}^{(I)}\left(\pi_{B|A}^{(U)}-\pi_{A|A}^{(I)}\right) \\
k^{(U)}(k^{(U)}-1)q_{B|A}^{(U)}q_{A|A}^{(U)}\left(\pi_{B|A}^{(U)}-\pi_{A|A}^{(U)}\right)
}\right)
\right] \\
&+O(\delta^2), 
\end{split}
\end{equation}
\begin{equation} \label{EpAAI}
\begin{split}
\dot{p}_{AA}^{(I)}=
&\sum_{k_A^{(I)},k_A^{(O)},k_A^{(U)}}\frac{k_A^{(I)}+k_A^{(O)}}{k^{(I)}N}\text{Prob}\left(\Delta p_{AA}^{(I)}=\frac{k_A^{(I)}+k_A^{(O)}}{k^{(I)}N}\right) \\
&+\sum_{k_A^{(I)},k_A^{(O)},k_A^{(U)}}\left(-\frac{k_A^{(I)}+k_A^{(O)}}{k^{(I)}N}\right)\text{Prob}\left(\Delta p_{AA}^{(I)}=-\frac{k_A^{(I)}+k_A^{(O)}}{k^{(I)}N}\right) \\
=&\frac{1}{N\left(k^{(I)}+k^{(U)}\right)}\left\{\left[\left(2k^{(I)}-1\right)\left(q_{A|B}^{(I)}-q_{A|A}^{(I)}\right)+1\right]p_{AB}^{(I)}+2k^{(U)}\left(q_{A|B}^{(I)}-q_{A|A}^{(I)}\right)p_{AB}^{(U)}\right\}+O(\delta),
\end{split}
\end{equation}
and 
\begin{equation} \label{EpAAU}
\begin{split}
\dot{p}_{AA}^{(U)}=
&\sum_{k_A^{(I)},k_A^{(O)},k_A^{(U)}}\frac{2k_A^{(U)}}{k^{(U)}N}\text{Prob}\left(\Delta p_{AA}^{(I)}=\frac{k_A^{(I)}+k_A^{(O)}}{k^{(I)}N}\right) \\
&+\sum_{k_A^{(I)},k_A^{(O)},k_A^{(U)}}\left(-\frac{2k_A^{(I)}}{k^{(U)}N}\right)\text{Prob}\left(\Delta p_{AA}^{(I)}=-\frac{k_A^{(I)}+k_A^{(O)}}{k^{(I)}N}\right) \\
=&\frac{2}{N\left(k^{(I)}+k^{(U)}\right)}\left\{\left[\left(k^{(U)}-1\right)\left(q_{A|B}^{(U)}-q_{A|A}^{(U)}\right)+1\right]p_{AB}^{(U)}+k^{(I)}\left(q_{A|B}^{(U)}-q_{A|A}^{(U)}\right)p_{AB}^{(I)}\right\}+O(\delta).
\end{split}
\end{equation}
Analyzing Eqs.~(\ref{EpA}-\ref{EpAAU}), for sufficiently small selection strength $\delta$, $p_{AA}^{(I)}$ and $p_{AA}^{(U)}$ reaches the equilibrium much faster than $p_{A}$.
The equilibrium can be obtained by solving  
\begin{equation}
\begin{split}
& \left[\left(2k^{(I)}-1\right)\left(q_{A|B}^{(I)}-q_{A|A}^{(I)}\right)+1\right]p_{AB}^{(I)}+2k^{(U)}\left(q_{A|B}^{(I)}-q_{A|A}^{(I)}\right)p_{AB}^{(U)}=0, \\
& \left[\left(k^{(U)}-1\right)\left(q_{A|B}^{(U)}-q_{A|A}^{(U)}\right)+1\right]p_{AB}^{(U)}+k^{(I)}\left(q_{A|B}^{(U)}-q_{A|A}^{(U)}\right)p_{AB}^{(I)}=0. \label{time_scale1}
\end{split}
\end{equation}
Let $x=q_{A|A}^{(I)}-q_{A|B}^{(I)}$ and $y=q_{A|A}^{(U)}-q_{A|B}^{(U)}$, using $q_{A|B}^{(Z)}-q_{A|A}^{(Z)}=\left(p_A-q_{A|A}^{(Z)}\right)/(1-p_A)$, we have 
\begin{equation}
\begin{split}
q_{A|A}^{(I)} =& (1-x)p_A+x, \\
q_{A|A}^{(U)} =& (1-y)p_A+y.
\end{split}\label{qAA}
\end{equation}

~\\
In the case of  $k^{(I)}=0$, from Eq.~(\ref{time_scale1}), we easily get $x=0$ and $y=1/\left(k^{(U)}-1\right)$, in agreement with prior work in the bi-directed setting \cite{2006-Ohtsuki-p502-505}.
For $k^{(U)}=0$, we have $x=1/\left(2k^{(I)}-1\right)$ and $y=0$.
In the following, we focus on the case with $k^{(I)}\ge 1,  k^{(U)}\ge 1$.
Moving the second terms in Eq.~(\ref{time_scale1}) to the right side and multiplying the two equations, we have 
\begin{equation}
y=\frac{-(2k^{(I)}-1)x+1}{(2k^{(I)}+k^{(U)}-1)x+k^{(U)}-1}. \label{yx}
\end{equation}
Substituting $p_{AB}^{(I)}=p_A\left(1-q_{A|A}^{(I)}\right)$ and  $p_{AB}^{(U)}=p_A\left(1-q_{A|A}^{(U)}\right)$  into the first equality in Eq.~(\ref{time_scale1}), and using Eqs.~(\ref{qAA},\ref{yx}), we have 
\begin{equation}
ax^3+bx^2+cx+d=0, \label{cubic}
\end{equation}
where 
\begin{equation}
\begin{split}
a=& 4\left(k^{(I)}\right)^2+2k^{(I)}k^{(U)}-4k^{(I)}-k^{(U)}+1, \\
b=& -4\left(k^{(I)}\right)^2-8k^{(I)}k^{(U)}-2\left(k^{(U)}\right)^2+3k^{(U)}+1,\\
c=& -2k^{(I)}k^{(U)}+4k^{(I)}-2\left(k^{(U)}\right)^2+5k^{(U)}-1,\\
d=& k^{(U)}-1.
\end{split}
\end{equation}
According to Eq.~(\ref{time_scale1}), the solution of Eq.~(\ref{cubic}), $x^*$, must satisfy $-1< x^*< 1$ and $-1< y(x^*)< 1$, which gives $ -\left(k^{(U)}-2\right)/\left(4k^{(I)}+k^{(U)}-2\right)< x^*<1 $.
Defining $G(x) = ax^3+bx^2+cx+d$, with $a>0$, we have 
\begin{equation}
G(-\infty)<0, \quad\quad G\left(-\frac{k^{(U)}-2}{4k^{(I)}+k^{(U)}-2}\right)>0, \quad\quad G(1)<0, \quad\quad G(\infty)>0.
\end{equation}
Therefore, Eq.~(\ref{cubic}) has and has only one solution in the interval  $(-\left(k^{(U)}-2\right)/\left(4k^{(I)}+k^{(U)}-2\right),1)$.
The other two respectively lie in 
$(-\infty,-\left(k^{(U)}-2\right)/\left(4k^{(I)}+k^{(U)}-2\right))$ and $(1,\infty)$.

~\\
Defining quantities $m=(3ac-b^2)/(3a^2)$ and $n=(2b^3-9abc+27a^2d)/(27a^3)$, 
the three roots of Eq.~(\ref{cubic}) are described by
\begin{equation} \label{eq:cubic_solution}
x^*_i=-\frac{b}{3a}+2\sqrt{-\frac{m}{3}}\cos\left(\frac{1}{3}\arccos\left(\frac{3n}{2m}\sqrt{-\frac{3}{m}}\right)-i\frac{2\pi}{3}\right), 
\end{equation}
for $i=0,1,2$.
Since $x^*_0>x^*_1>x^*_2$, the solution for Eq.~(\ref{time_scale1}) is $x^*_1$.
Eq.~(\ref{cubic}) tells that $x^*_1$ only depends on the structure properties $k^{(I)}$ and $k^{(U)}$, while is independent of $p_A$. 
Then $y$ is obtained by substituting $x^*_1$ into Eq.~(\ref{yx}).
For simplicity, we still use $x$ to denote $x^*_1$.


\subsection{Diffusion process}
After obtaining $x$ and $y$, besides Eq.~(\ref{qAA}), we have  
\begin{equation}
\begin{split}
q_{B|A}^{(I)} &= (1-x)(1-p_A), \quad q_{A|B}^{(I)} = (1-x)p_A, \quad q_{B|B}^{(I)} = 1-(1-x)p_A,   \\
q_{B|A}^{(U)} &= (1-y)(1-p_A), \quad q_{A|B}^{(U)} = (1-y)p_A, \quad q_{B|B}^{(U)} = 1-(1-y)p_A.   
\end{split} \label{qBA}
\end{equation}
Substituting Eqs.~(\ref{qAA}) and ~(\ref{qBA}) into Eq.~(\ref{EpA}) gives 
\begin{equation}
\begin{split}
\text{E}(p_A)&=\frac{\delta}{N\left(k^{(I)}+k^{(U)}\right)^2}p_A(1-p_A)\left(\alpha p_A+\beta\right) \quad (\equiv\delta mp_A(1-p_A)\left(\alpha p_A+\beta\right)), \\
\text{Var}(p_A)&=\frac{2}{N^2\left(k^{(I)}+k^{(U)}\right)}p_A(1-p_A)\lambda \quad (\equiv np_A(1-p_A)),
\end{split}
\end{equation}
where 
\begin{equation} \label{eq:alpha_beta}
\begin{split}
\alpha =& \left[\lambda\gamma-(1-x)\mu-(1-y)\nu\right](a_1-b_1-c_1+d_1) \\
&+\left[\lambda\gamma-(1-y)\nu\right](a_2-b_2-c_2+d_2),\\
\beta =& \left[\left(k^{(I)}x+k^{(U)}y\right)\gamma-\mu x-\nu y\right]a_1
+(\lambda\gamma+\mu x+\nu y)b_1
-(\mu+\nu)c_1\\
&-(k^{(I)}+k^{(U)}-1)(\gamma+\lambda)d_1
+\left[\left(k^{(I)}x+k^{(U)}y\right)\gamma-\nu y\right]a_2 \\
&-\nu b_2
+(\lambda \gamma+\nu y)c_2
-\left[\left(k^{(I)}+k^{(U)}\right)\gamma-\nu\right]d_2, \\
\gamma =& k^{(I)}\left(k^{(I)}-1\right)(1-x^2)+k^{(U)}\left(k^{(U)}-1\right)(1-y^2)+2k^{(I)}k^{(U)}(1-xy),\\
\lambda =& k^{(I)}(1-x)+k^{(U)}(1-y),\\
\mu =& k^{(I)}(1-x)\left(k^{(I)}x+k^{(U)}y-x\right),\\
\nu =& k^{(U)}(1-y)\left(k^{(I)}x+k^{(U)}y-y\right).
\end{split}
\end{equation}

~\\
Introducing 
\begin{equation}
\psi(v) = \text{exp}\left(-\int^v \frac{2\text{E}(r)}{\text{Var}(r)}dr \right),
\end{equation}
we have the fixation probability $\rho_A(u)$, the probability that a proportion $u$ of $A$-players take over the whole population, given by 
\begin{equation}
\begin{split}
\rho_A(u)&=\frac{\int_0^u \psi(y) dy}{\int_0^1 \psi(y) dy}  =u+\frac{\delta m}{3n} u(1-u)\left[\alpha u+ (\alpha+3\beta)\right].
\end{split}
\end{equation} 

\subsection{Fixation probability}
Selection produces $\rho_A(1/N)>1/N$, the fixation probability under neutral drift, provided $\alpha+3\beta>0$.
Selection produces $\rho_A(1/N)>\rho_B(1/N)$, the fixation probability of B-players, provided $\alpha+2\beta>0$, which is equivalent to 
\begin{equation}
\begin{split}
& \left[\bar{\lambda}\gamma-(1+x)\mu-(1+y)\nu\right]a_1+\left[\lambda\gamma+(1+x)\mu+(1+y)\nu\right]b_1 \\ 
-&\left[\lambda\gamma+(1+x)\mu+(1+y)\nu\right]c_1-\left[\bar{\lambda}\gamma-(1+x)\mu-(1+y)\nu\right]d_1 \\
+& \left[\bar{\lambda}\gamma-(1+y)\nu\right]a_2-\left[\lambda\gamma+(1+y)\nu\right]b_2 \\ 
+&\left[\lambda\gamma+(1+y)\nu\right]c_2-\left[\bar{\lambda}\gamma-(1+y)\nu\right]d_2 > 0,
\end{split}  \label{general_formula}
\end{equation}
where 
\begin{equation}
\bar\lambda = k^{(I)}(1+x)+k^{(U)}(1+y).
\end{equation}
For payoff structure 
\begin{equation}
\bordermatrix{
  & \textit{A} & \textit{B} \cr
\textit{A} & a_1,a_2 & b_1,b_2 \cr
\textit{B} & c_1,c_2 & d_1,d_2 \cr
}
\rightarrow 
\bordermatrix{
  & \textit{A} & \textit{B} \cr
\textit{A} & -c,b & -c,b \cr
\textit{B} & 0,0 & 0,0 \cr
},
\end{equation}
the directions of donating action and strategy dispersal are identical, corresponding to the downstream case.
We have 
\begin{equation}
\rho_A>\rho_B \leftrightarrow \frac{b}{c}>\frac{\left(k^{(I)}+k^{(U)}\right)\gamma}{\left(k^{(I)}x+k^{(U)}y\right)\gamma-(1+y)\nu},\label{formula_case1}
\end{equation}
where $x$ is obtained from Eq.~(\ref{eq:cubic_solution}), $y$ from Eq.~(\ref{yx}), $\gamma$ and $\nu$ from Eq.~(\ref{eq:alpha_beta}).
For payoff structure 
\begin{equation}
\bordermatrix{
  & \textit{A} & \textit{B} \cr
\textit{A} & a_1,a_2 & b_1,b_2 \cr
\textit{B} & c_1,c_2 & d_1,d_2 \cr
}
\rightarrow 
\bordermatrix{
  & \textit{A} & \textit{B} \cr
\textit{A} & b,-c & 0,0 \cr
\textit{B} & b,-c & 0,0 \cr
},
\end{equation}
the directions of donating action and strategy dispersal are opposite, corresponding to the upstream case.
We have 
\begin{equation}
\rho_A>\rho_B \leftrightarrow \frac{b}{c}>\frac{\left(k^{(I)}+k^{(U)}\right)\gamma}{\left(k^{(I)}x+k^{(U)}y\right)\gamma-(1+x)\mu-(1+y)\nu}.
 \label{formula_case2}
\end{equation}

\subsection{Case of $k\gg 1$}
Let $p$ denote the fraction of directed edges and therefore $1-p$ the fraction of bidirected edges.
We have $k^{(I)}=k^{(O)}=kp/2$ and $k^{(U)}=k(1-p)$.
Defining $\widetilde{x}=kx$ and substituting these into Eq.~(\ref{cubic}), we have 
\begin{equation}
\widetilde{a}\widetilde{x}^3+\widetilde{b}\widetilde{x}^2+\widetilde{c}\widetilde{x}+\widetilde{d}=0, \label{cubic_k}
\end{equation}
where 
\begin{equation}
\begin{split}
\widetilde{a}=& pk^2-(1+p)k+1, \\
\widetilde{b}=& (p^2-2)k^3+3(1-p)k^2+k,\\
\widetilde{c}=& (-p^2+3p-2)k^4+(5-3p)k^3-k^2,\\
\widetilde{d}=& (1-p)k^4-k^3.
\end{split}
\end{equation}
For $k\gg 1$, a solution for Eq.~(\ref{cubic_k}) is 
\begin{equation}
\widetilde{x}^*=\frac{1}{2-p},
\end{equation}
which is the root for Eq.~(\ref{time_scale1}).
Then substituting $x=1/((2-p)k)$ into Eq.~(\ref{general_formula}), and replacing $k^{(I)}$, $k^{(U)}$ with $kp/2$, $k(1-p)$, we can rewrite Eq.~(\ref{general_formula}) to be 
\begin{equation}
\begin{split}
&\frac{1}{8}\left[(2-p)^3k^3-2(2-p)(3-p)k^2\right](a_1-d_1+a_2-d_2) \\
+&\frac{1}{8}\left[(2-p)^3k^3-2(2-p)(7-4p)k^2\right](b_1-c_1-b_2+c_2)+o(k^2)>0,
\end{split}
\end{equation}
which can further simplified to be 
\begin{equation}
\left[(2-p)^2k-6+2p\right](a_1-d_1+a_2-d_2) \\
+\left[(2-p)^2k-14+8p\right](b_1-c_1-b_2+c_2)>0.
\end{equation}
Accordingly, both critical benefit-to-cost ratios in Eqs.~(\ref{formula_case1}) and ~(\ref{formula_case2}) are 
\begin{equation}
\left(\frac{b}{c}\right)^*=\frac{(2-p)^2}{4-3p}k,
\end{equation} 
where $\left(b/c\right)^*$ is minimized when $p=2/3$, and the minimal threshold is $\left(b/c\right)^*_{\text{min}}=8k/9$.

\subsection{Any directed network}
We refer to a prior work \cite{McAvoy2021} to provide the condition for the evolution of cooperation in general directed networks, with independent structures for interaction and strategy dispersal.
Let $w_{ij}^{[1]}$ (resp. $w_{ij}^{[2]}$) denote the edge weight in the interaction (resp. dispersal) network.
Let $p_{ij}=w_{ji}^{[2]}/\sum_{\ell}w_{\ell i}^{[2]}$ and $\pi_i$ denote the probability that a mutant in node $i$ takes over the whole population under neutral drift.
We can obtain $\pi_i$ by solving $\sum_{i}\pi_i=1$ and $\pi_i=\sum_{j}p_{ji}\pi_j$.
Using theorem 1 in Ref. \cite{McAvoy2021}, we have 
\begin{align}
\eta_{ij} &= 
\begin{cases}
1+\frac{1}{2}\sum_{k\in\mathcal{N}}p_{ik}\eta_{kj}+\frac{1}{2}\sum_{k\in\mathcal{N}}p_{jk}\eta_{ik} & i\neq j , \\
& \\
0 & i=j .
\end{cases}\label{eq:eta_recurrence}
\end{align}
The critical benefit-to-cost ratio for $\rho_A>\rho_B$ under weak selection is given by 
\begin{equation}
\left(\frac{b}{c}\right)^*=\frac{v_2}{u_2-u_0}
\end{equation}
where 
\begin{equation}
\begin{split}
u_0 &= \sum_{i,j,\ell\in\mathcal{N}}\pi_ip_{ij}\left(-\eta_{j\ell}\right)w_{\ell j}^{[1]}, \quad \quad u_2 = \sum_{i,j,k,\ell\in\mathcal{N}}\pi_ip_{ij}p_{ik}\left(-\eta_{j\ell}\right)w_{\ell k}^{[1]}, \\
v_2 &= \sum_{i,j,k,\ell\in\mathcal{N}}\pi_ip_{ij}p_{ik}\left(-\eta_{jk}\right)w_{k\ell}^{[1]}.
\end{split}
\end{equation}
The `downstream' case corresponds to the structure with $w_{ij}^{[2]}=w_{ij}^{[1]}$ for any $i,j\in\mathcal{N}$.
The `upstream' case corresponds to the structure with $w_{ij}^{[2]}=w_{ji}^{[1]}$.

\section{Network motifs}
Let $k_i$ denote node $i$'s degree, including $k_i^{(I)}$ incoming edges, $k_i^{(O)}$ outgoing edges, and $k_i^{(U)}$ bidirected/undirected edges, i.e. $k_i=k_i^{(I)}+k_i^{(O)}+k_i^{(U)}$.
In the following, we treat each bidirected/undirected as two directed edges, namely an `incoming' and an `outgoing' edges.
In the downstream dispersal, we consider the motif of triangular cycles, such as $i\rightarrow j\rightarrow \ell\rightarrow i$ ($j\ne \ell$). 
For node $i$, the number of such triangular cycles is $\sum_{j,\ell}w_{ij}w_{j\ell}w_{\ell i}$.
For node $i$, the number of in-out pairs (i.e. an incoming edge and an outgoing edge, like $y\rightarrow i$ and $i\rightarrow j$ but $y\ne j$) is $k_i^{(I)}k_i^{(O)}+k_i^{(U)}(k_i-1)$.
For each in-out pair, if there exists an edge from the target node of $i$'s outgoing edge to the source node of $i$'s incoming edge, a triangular cycle appears. 
Therefore, the number of in-out pairs is the possibly largest number of triangular cycles for node $i$.
We introduce a quantity
\begin{equation}
\mathcal{C}_1=\frac{\sum_{i,j,\ell\in\mathcal{N}} w_{ij}w_{j\ell}w_{\ell i}}{\sum_{i\in\mathcal{N}} \left(k_i^{(I)}k_i^{(O)}+k_i^{(U)}(k_i-1)\right)}
\end{equation}
to measure the global frequency of triangular cycles in a directed network.
A larger $\mathcal{C}_1$ means more triangular cycles.

~\\
In the case of upstream dispersal, we consider the motif of in-in pairs, such as $j\rightarrow i \leftarrow \ell$ ($j\ne \ell$).
For node $i$, the number of in-in pairs is $\left(k_i^{(I)}+k_i^{(U)}\right)\left(k_i^{(I)}+k_i^{(U)}-1\right)$, and the number of all edge pairs is 
$\left(k_i+k_i^{(U)}\right)\left(k_i+k_i^{(U)}-1\right)$.
We introduce the quantity 
\begin{equation}
\mathcal{C}_2=\frac{\sum_{i\in\mathcal{N}} \left(k_i^{(I)}+k_i^{(U)}\right)\left(k_i^{(I)}+k_i^{(U)}-1\right)}{\sum_{i\in\mathcal{N}} \left(k_i+k_i^{(U)}\right)\left(k_i+k_i^{(U)}-1\right)}.
\end{equation}
This quantity measures the normalized fraction of in-in pairs in the directed network, and a larger value of $\mathcal{C}_2$ means a greater frequency of in-in pairs.

\begin{figure*}[!h]
\centering
\includegraphics[width=1\textwidth]{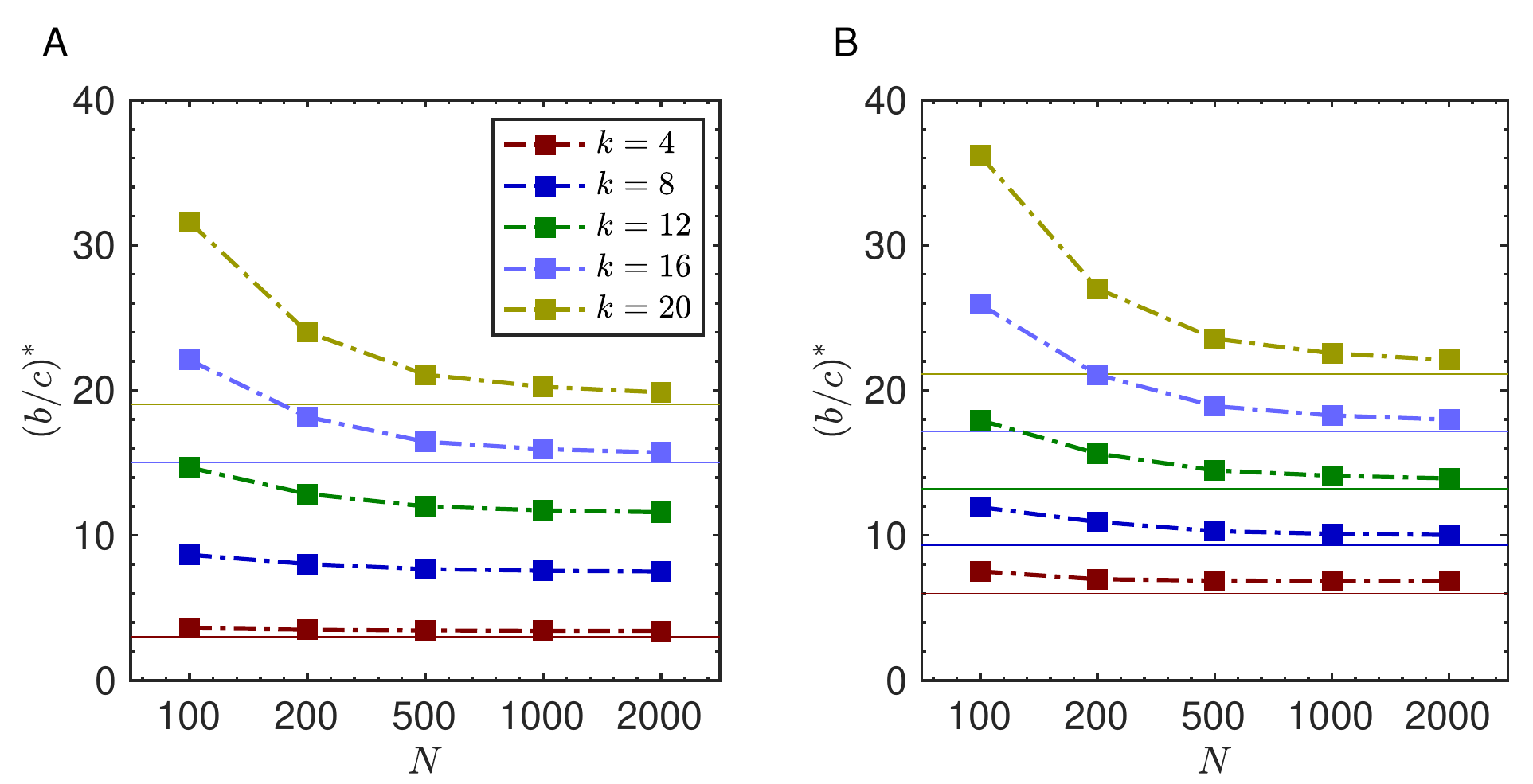}
\caption{\label{sfig:1} \textbf{Exact results (dots) approach those predicted by the pair approximation (horizontal lines) for large population size $N$}.
(A) downstream dispersal. 
(B) upstream dispersal.
}
\end{figure*}

\begin{figure*}[!h]
\centering
\includegraphics[width=1\textwidth]{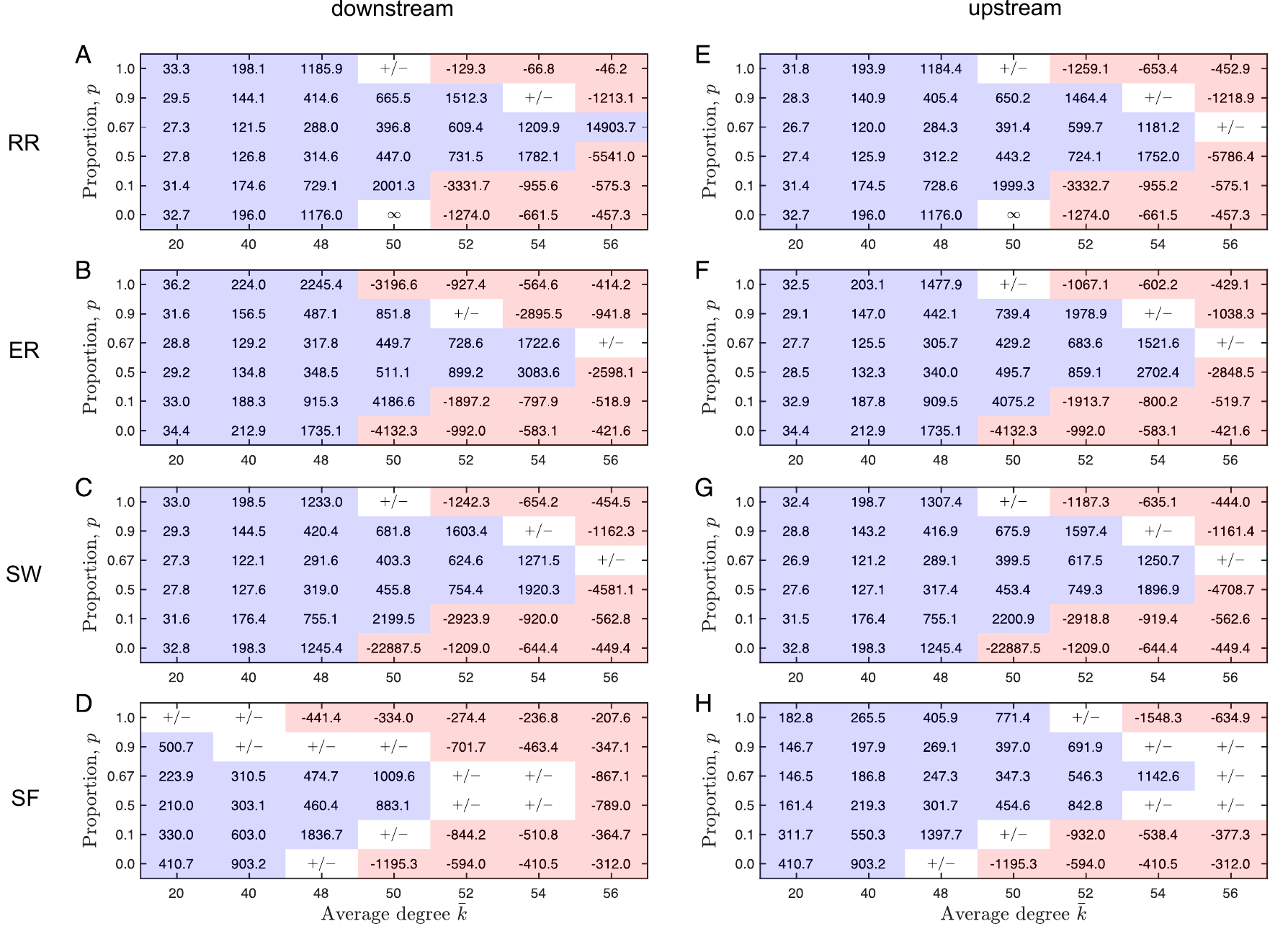}
\caption{\label{sfig:2} \textbf{Intermediate frequency of directed edges is optimal for cooperation on random networks}.
We consider four classes of networks:  random regular networks (RR), Erd\"{o}s-R\'{e}nyi networks (ER) \cite{1960-Erdoes-p17-61}, Watts-Strogatz small-world networks (SW) with rewiring probability $0.1$ \cite{1998-Watts-p440-442}, and Barab\'{a}si-Albert scale-free networks (BA-SF) \cite{1999-Barabasi-p509-512}.
For each class, we generated $10,000$ undirected networks each with $N=100$ nodes, for several values of the average node degree  $\bar{k}$. For each such network we randomly select a proportion $p$ of edges and converted them to be uni-directional, with randomly chosen orientation. For each resulting network we compute the critical benefit-to-cost ratio $(b/c)^*$ required to favor cooperation, and we plot the mean value, given $p$ and $\bar{k}$, in the tables.
In the regions displayed in blue, all ratios are positive and cooperation can evolve for some choice of benefits and costs.
In red regions, all ratios are negative and spite is favored instead of cooperation.
The symbol `$+/-$' means that a fraction of critical ratios are positive and a fraction negative; and the symbol `$\infty$' means that cooperation is never favored, regardless of how large the benefit.
}
\end{figure*}

\begin{figure*}[!h]
\centering
\includegraphics[width=0.5\textwidth]{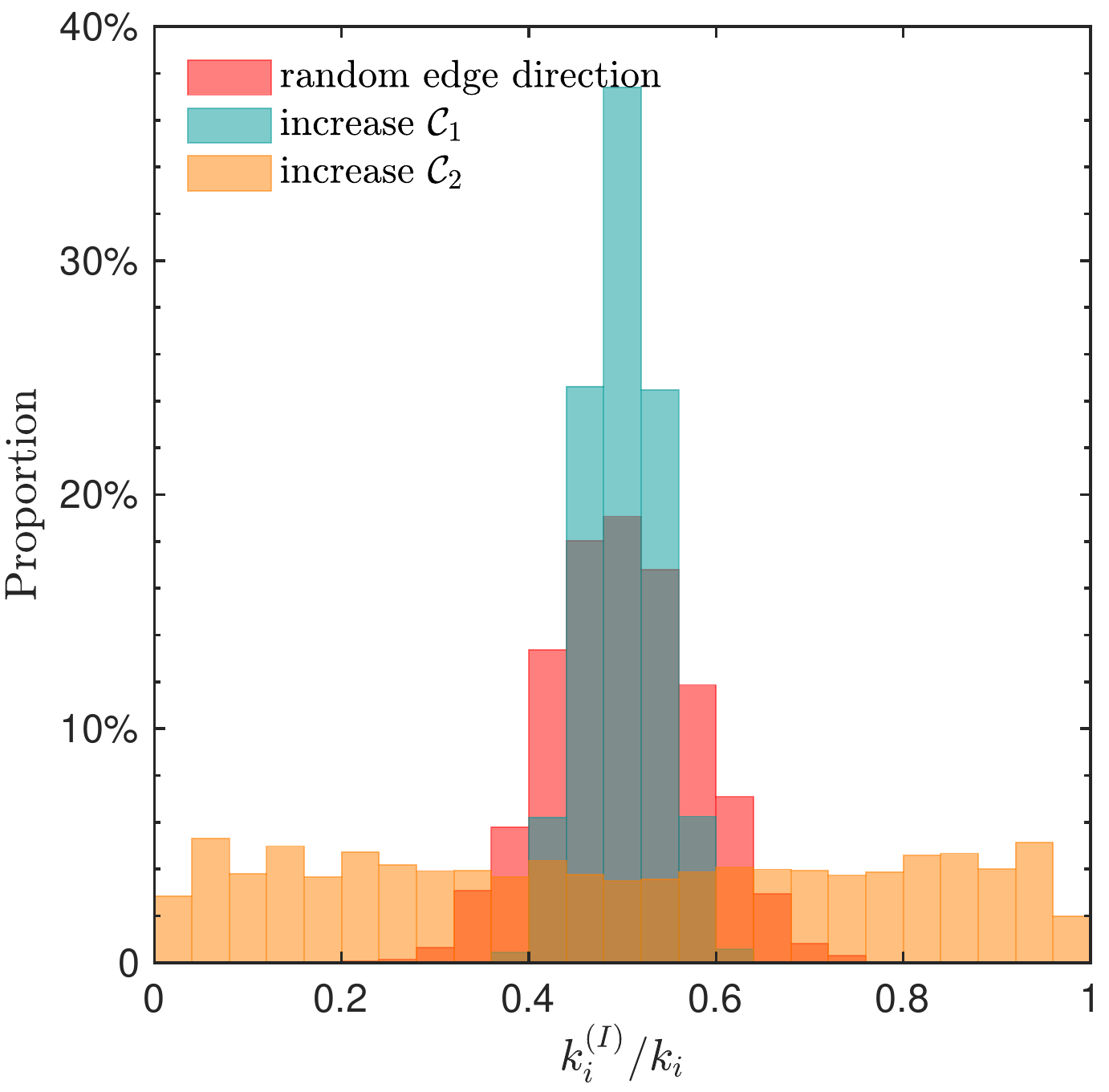}
\caption{\label{sfig:3} \textbf{The distribution of node in-degrees for different motif frequencies}.
Random edge orientation leads to an intermediate heterogeneity of the node in-degree distribution.
Increasing the frequency of triangular cycles ($\mathcal{C}_1$) leads to the homogeneity in node in-degree distribution, namely each node has roughly the same number of incoming neighbors as outgoing neighbors.
Increasing the frequency in-in pairs ($\mathcal{C}_2$) causes large heterogeneity in node in-degree distribution, namely some nodes have many more incoming neighbors than outgoing neighbors, while other nodes have many more outgoing neighbors than incoming neighbors.
}
\end{figure*}

\begin{figure*}[!h]
\centering
\includegraphics[width=0.7\textwidth]{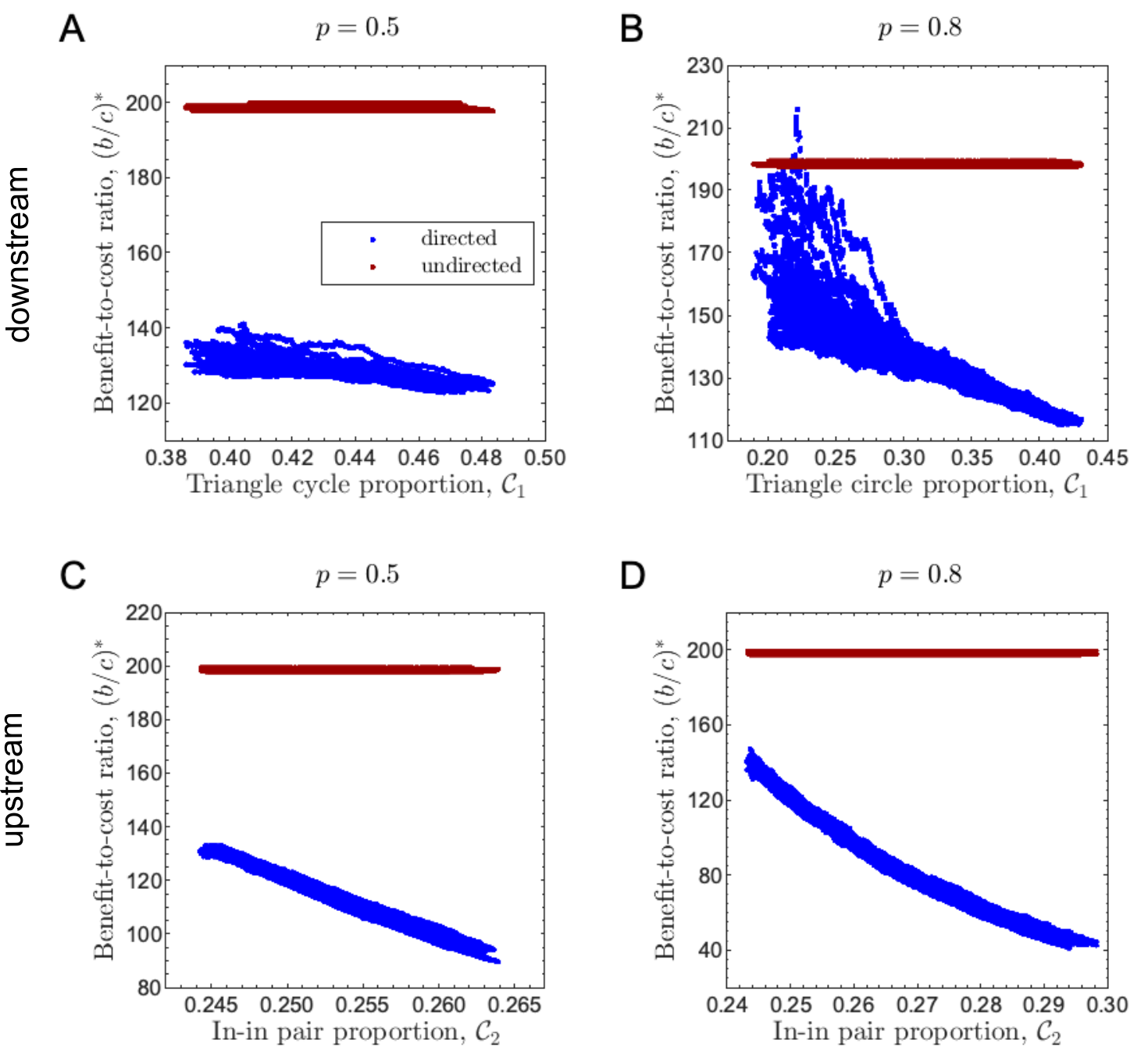}
\caption{\label{sfig:4} \textbf{Adjusting edge orientation effects evolutionary outcomes in semi-directed networks.}
Here we investigate small-world networks with $N=100$ and $\bar{k}=40$, and with fraction of directed edges $p=0.5$ (AC) or $p=0.8$ (BD) (see Figure~\ref{fig:4} in the main text for all other parameters and detailed caption). 
}
\end{figure*}

\begin{figure*}[!h]
\centering
\includegraphics[width=0.7\textwidth]{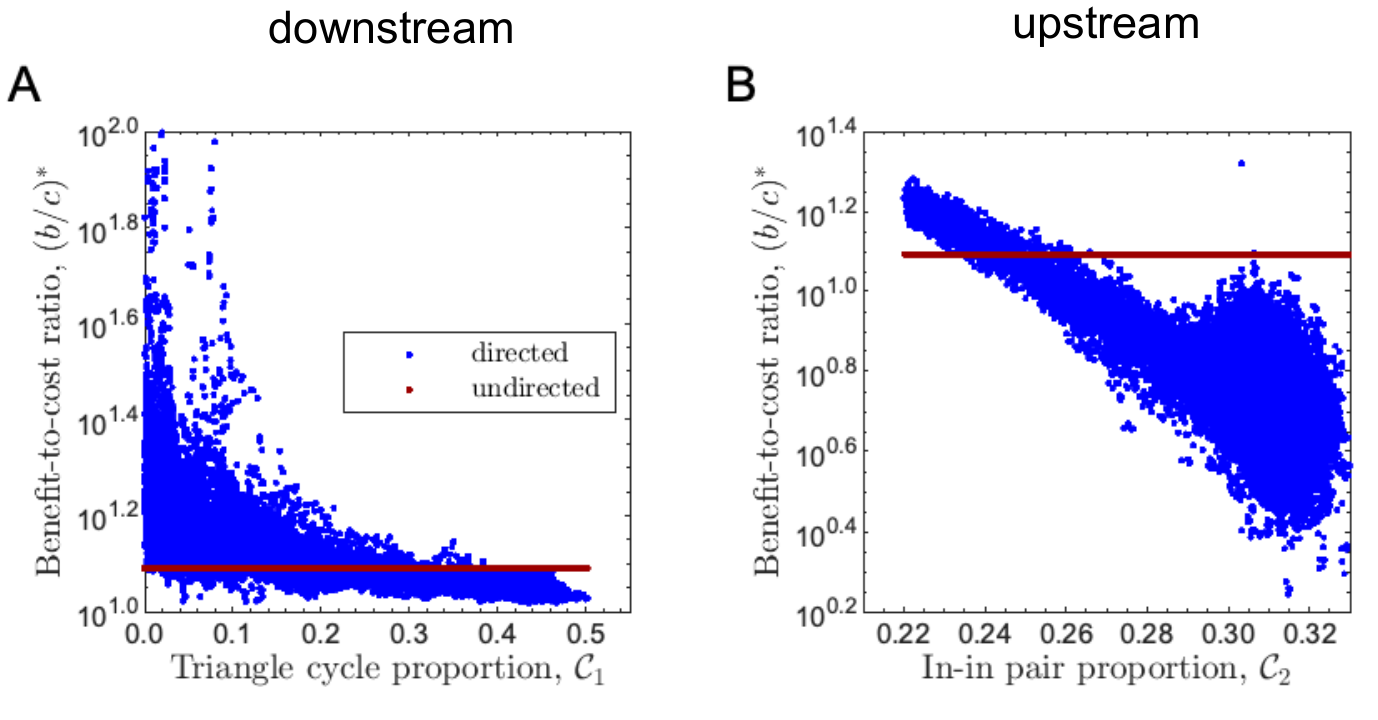}
\caption{\label{sfig:5} \textbf{Adjusting edge orientation effects evolutionary outcomes in sparse directed networks.}
Here we investigate small-world networks with $N=100$ and $\bar{k}=10$, $p=1.0$ (see Figure~\ref{fig:4} in the main text for all other parameters and detailed caption). 
}
\end{figure*}

\begin{figure*}[!h]
\centering
\includegraphics[width=0.7\textwidth]{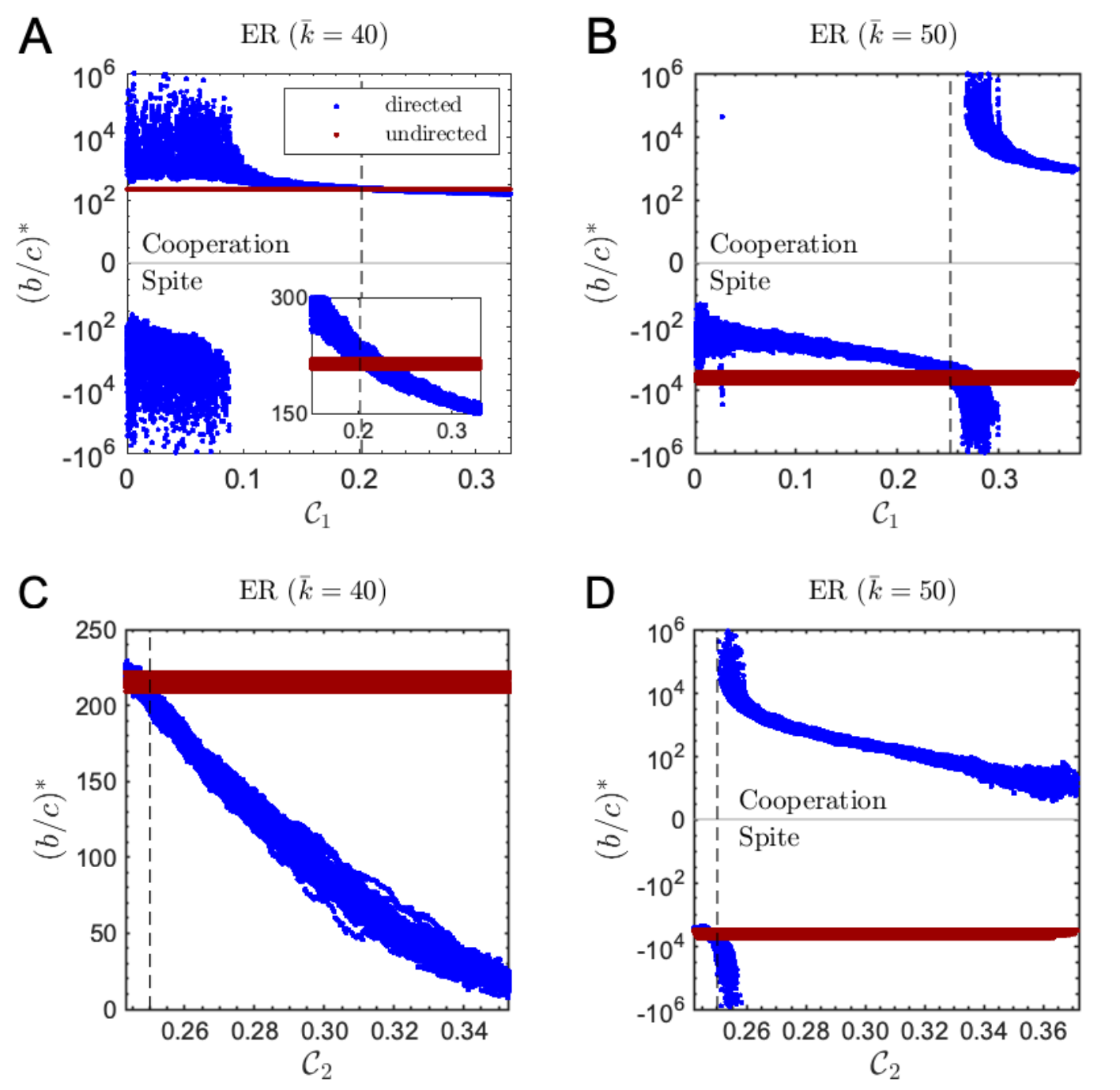}
\caption{\label{sfig:6} \textbf{Adjusting edge orientation effects evolutionary outcomes in random networks.}
Here we investigate ER random networks (see Figure~\ref{fig:4} in the main text for all other parameters and detailed caption) .
}
\end{figure*}


\end{document}